\documentclass[12pt]{iopart}
\pdfoutput=1
\expandafter\let\csname equation*\endcsname\relax
\expandafter\let\csname endequation*\endcsname\relax

\usepackage{color}
\usepackage[breakwords]{truncate}
\usepackage[font={small}]{caption}
\usepackage{physics}
\usepackage{graphicx}
\usepackage{cite} 
\usepackage{subcaption}
\usepackage{etoolbox}
\usepackage[section]{placeins}
\usepackage[sectionbib]{chapterbib}
\usepackage{amsfonts}
\usepackage[bookmarks,bookmarksopen,bookmarksdepth=2]{hyperref}

\makeatletter
\newrobustcmd{\fixappendix}{%
  \patchcmd{\l@section}{1.5em}{7em}{}{}%
  \patchcmd{\l@subsection}{2.3em}{7em}{}{}%
}
\makeatother

\appto\appendix{
\addtocontents{toc}{\fixappendix}
\addtocontents{toc}{\protect\setcounter{tocdepth}{1}}}

\begin{document}
\newcommand{\ben}[1]{\textcolor{blue}{\textbf{#1}}}

\title{Expected maximum of bridge random walks \& L\'evy flights}
\author{Benjamin De Bruyne}
\address{LPTMS, CNRS, Univ.\ Paris-Sud, Universit\'e Paris-Saclay, 91405 Orsay, France}
\author{Satya N. Majumdar}
\address{LPTMS, CNRS, Univ.\ Paris-Sud, Universit\'e Paris-Saclay, 91405 Orsay, France}
\author{Gr{\'e}gory Schehr }
\address{Sorbonne Universit\'e, Laboratoire de Physique Th\'eorique et Hautes Energies, CNRS UMR 7589, 4 Place Jussieu, 75252 Paris Cedex 05, France}
\eads{\mailto{benjamin.debruyne@centraliens.net}, \mailto{satya.majumdar@universite-paris-saclay.fr},
\mailto{gregory.schehr@u-psud.fr}}
\begin{abstract}
We consider one-dimensional discrete-time random walks (RWs) with arbitrary symmetric and continuous jump distributions $f(\eta)$, including the case of L\'evy flights. We study the expected maximum ${\mathbb E}[M_n]$ of bridge RWs, i.e., RWs starting and ending at the origin 
after $n$ steps. We obtain an exact analytical expression for ${\mathbb E}[M_n]$ valid for any $n$ and jump distribution $f(\eta)$, which we then analyze in the large $n$ limit up to second leading order term. For jump distributions whose Fourier 
transform behaves, for small $k$, as $\hat f(k) \sim 1 - |a\, k|^\mu$ with a L\'evy index $0<\mu \leq 2$ and an arbitrary length scale $a>0$,  
we find that, at leading order for large $n$, ${\mathbb E}[M_n]\sim a\, h_1(\mu)\, n^{1/\mu}$. We obtain an explicit expression for the amplitude $h_1(\mu)$ and find that it carries the signature of the bridge condition, being different from its counterpart for the free random walk. For $\mu=2$, we find that the second leading order term is a constant, which, quite remarkably, is the same as its counterpart for the free RW. For generic $0< \mu < 2$, this second leading order term is a growing function of $n$, which depends non-trivially on further details  of $\hat f (k)$, beyond the L\'evy index $\mu$. 
Finally, we apply our results to compute the mean perimeter of the convex hull of the $2d$ Rouse polymer chain and of the $2d$ run-and-tumble particle, as well as to the computation of the survival probability in  
a bridge version of the well-known ``lamb-lion'' capture problem.
\end{abstract}
\newpage
{\pagestyle{plain}
 \tableofcontents
\cleardoublepage}

\section{Introduction and main results}
\subsection{Introduction}
Brownian motion has shown to be a successful model to describe a tremendous amount of natural phenomena ranging from biology \cite{Koshland80} to astrophysics \cite{Chandrasekhar43}. Brownian motion is a stochastic process that is continuous in both space and time. In some practical situations, the continuous-time limit underlying Brownian motion is not applicable due to the discreteness of the problem and one needs to consider discrete-time random walks (RWs) \cite{Montroll84,BouchaudAn90}. This is for instance the case in combinatorial problems in computer science \cite{Majumdar05Ein,KnuthThe98} or in the statistics of polymer chains in material science \cite{Rouse53}. In its simplest form, a one-dimensional discrete-time RW $x_m$ is described by a Markov rule
\begin{align}
  x_{m+1} &= x_{m} + \eta_{m}\,, \label{eq:XmI}
\end{align}
starting from $x_0$, where $\eta_m$'s  are independent and identically distributed (i.i.d.) random variables drawn from a symmetric and continuous jump distribution $f(\eta)$. Discrete-time RWs with finite variance jump distributions converge to Brownian motion in the limit $n\rightarrow \infty$ due to the central limit theorem. When the variance is not finite, they converge to L\'evy processes and are generally much harder to study. While Brownian motion and L\'evy processes are interesting by themselves, RWs are also important to study, especially in the limit of large but finite $n$ where they display finite size effects that are lost in the $n\rightarrow\infty$ limit.

Finite size effects appear for instance in the study of extreme value statistics of discrete-time RWs \cite{ComtetPrecise05}. For instance, let us consider the maximum $M_n$ of a discrete-time random walk after $n$ steps 
 \begin{align}
  M_n = \text{max}\{x_0,\ldots,x_n\}\,,\label{eq:maxdef}
\end{align}
an observable that is commonly studied in the mathematics literature \cite{Kac54,Darling56,Spitzer56,Doney87,Doney08,CD10,Kuznetsov11} as well as in natural and practical contexts, such as animal foraging where the spatial extent of their territory can be characterized by the extreme points of their trajectories \cite{Dumonteil13,Majumdar21convex,Randon09Convex,Majumdar10Random,GrebenkovConvex17,Schawe18Large,Cauchy32}. 
For a free random walk starting from the origin $x_0=0$, it is well known that the expected maximum is simply given by \cite{Kac54,Spitzer56}
\begin{align}
  {\mathbb E}[M_n]
  &=\sum_{m=1}^n \frac{1}{m}
   \int_{0}^\infty dy\, y\, P(y,m) \,,\label{eq:maxfree}
\end{align}
where ${\mathbb E}[\cdot]$ denotes an average over all possible RW trajectories in (\ref{eq:XmI}), and $P(y,m)$ is the forward propagator of the free random walk, which is the probability density that the position $y$ is reached in $m$ steps given that it started at the origin. It is simply given~by
 \begin{align}
   P(y,m) = \int_{-\infty}^\infty \frac{dk}{2\pi} \,\left[\hat f(k)\right]^m\,\rme^{i\,k\,y}\,,\label{eq:PI}
\end{align}
 where $\hat f(k)$ is the Fourier transform of the symmetric jump distribution $f(\eta)$. Finite size effects appear when considering the large but finite $n$ limit of the expected maximum ${\mathbb E}[M_n]$. For jump distributions with a finite variance $\sigma$, such that $\hat f(k) \sim 1 - \sigma^2\,k^2/2$ for small $k$, the expected maximum ${\mathbb E}[M_n]$ grows, to leading order for large $n$, as $\sqrt{2\,n\,\sigma^2/\pi}$, a result that can be easily obtained from the diffusive (i.e., Brownian) limit. In contrast, the second leading order term in the asymptotic limit of the expected maximum is non trivial to obtain and contains the leading finite size correction \cite{ComtetPrecise05}. This finite size correction is actually important as it appears as a leading order term in the thermodynamic limit of various geometrical properties such as the difference between the expected maximum ${\mathbb E}[M_n]$ and the average absolute position of the RW ${\mathbb E}[|x_n|]$ \cite{ComtetPrecise05}. In addition, it turns out that it also has applications in various algorithmic problems \cite{Coffman93,Coffman98}. This leading finite size correction was obtained in \cite{ComtetPrecise05} where it was shown that, for finite variance jump distributions (with additional regularity properties \cite{ComtetPrecise05}), this correction is a constant $\gamma$ such that
\begin{align}
  \frac{{\mathbb E}[M_n]}{\sigma}  \sim \sqrt{\frac{2\,n}{\pi}}+\gamma \,,\quad n\rightarrow \infty\,,\label{eq:mnfree}
\end{align}
where $\gamma$ is given by
\begin{align}
 \gamma =  \frac{1}{\pi\sqrt{2}}\,\int_0^\infty \frac{dk}{k^2}\,\ln\left[\frac{1-\hat f\left(\frac{\sqrt{2}}{\sigma}k\right)}{k^2}\right]\,,\label{eq:gamma}
\end{align}
where $\hat f(k)$ is the Fourier transform of the jump distribution. Interestingly, the constant $\gamma$ depends on the full details of the jump distribution and takes non-trivial values (e.g., for Gaussian jump distributions, $\gamma= \zeta(1/2)/\sqrt{2\pi}$ where $\zeta$ is the Riemann zeta function). For such distributions with finite variance, it was shown that the difference between the expected maximum ${\mathbb E}[M_n]$ and the average absolute value of the position of the walker ${\mathbb E}[|x_n|]$ tends to this constant \cite{ComtetPrecise05}
\begin{align}
 \lim_{n \to \infty}  \frac{{\mathbb E}[M_n]-{\mathbb E}[|x_n|]}{\sigma} = \gamma \,,\label{Mnxn}
\end{align}
which is not obvious and is a signature of discreteness of the process that remarkably persists in the limit $n \to \infty$.
 
\begin{figure}[t]
  \begin{center}
    \includegraphics[width=0.5\textwidth]{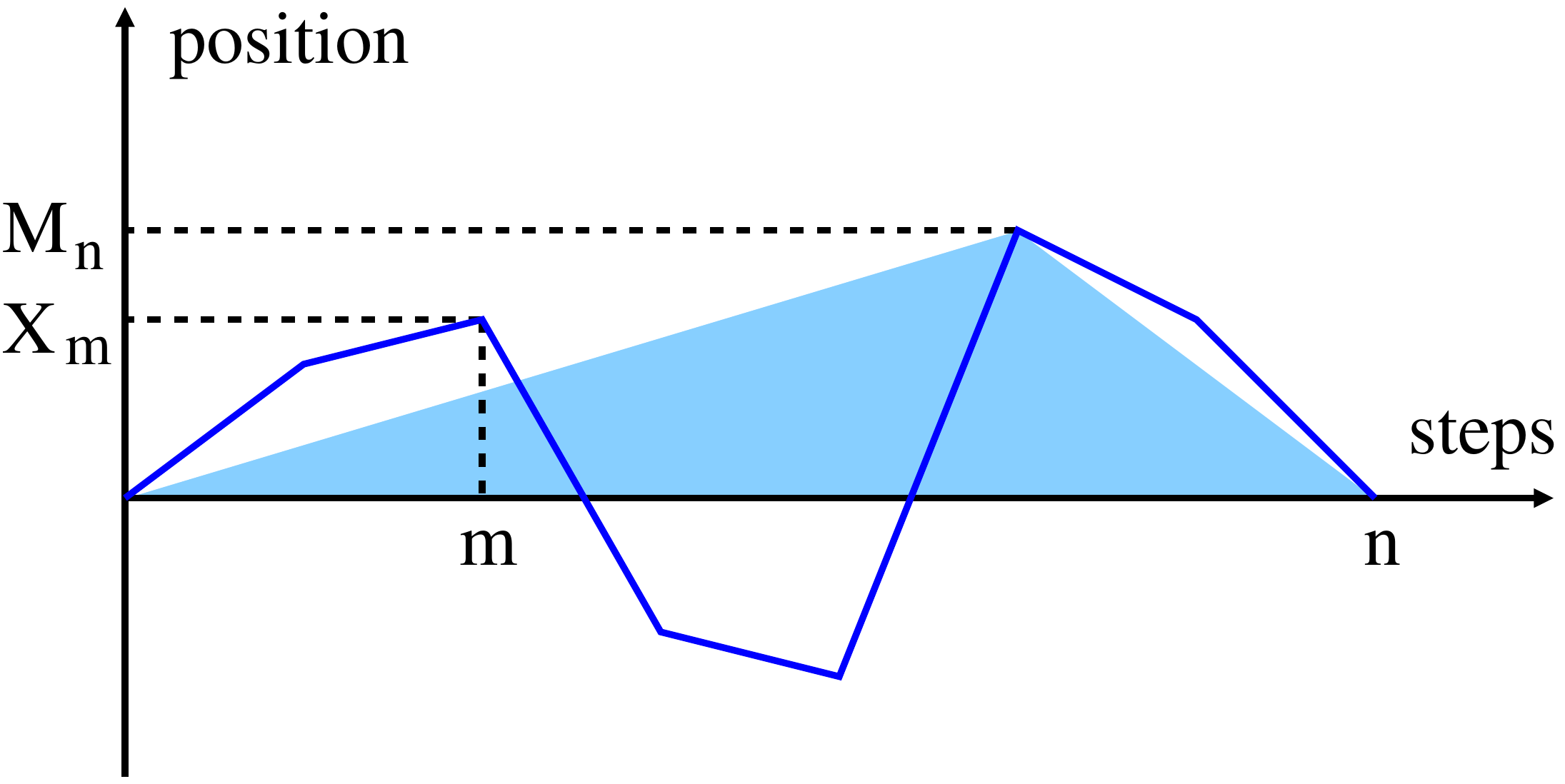}
    \caption{A bridge RW of $n$ steps is a RW that is constrained to start at the origin and return to the origin after $n$ steps. The maximum of a bridge RW of $n$ steps is denoted $M_n$. For finite variance jump distributions, where the L\'evy index $\mu=2$ in phase I (see figure \ref{fig:phaseDiagram}), the scaled difference between the expected area $n\,{\mathbb E}[M_n]/2$ of the triangle indicated in light blue and the expected absolute area under a bridge random walk $\sum_{m=0}^n {\mathbb E}[|X_m|]$ tends to the constant $\gamma/2$ where $\gamma$ is given in (\ref{eq:gamma}) and (\ref{eq:maxga}). }
    \label{fig:bridge}
  \end{center}
\end{figure}

 In addition to being discrete in time, some problems are subjected to some global constraints and are described by \emph{constrained} RWs. One prominent example is the case of bridge RWs which appear in several problems in computer science and graph theory \cite{Majumdar05Ein,KnuthThe98,FlajoletOn98,MajumdarExact02,HararyDyna97,Takacs91}. Bridge RWs also manifest themselves frequently in physics and mathematics such as in problems of fluctuating interfaces \cite{Majumdar04Flu1,Majumdar05Flu2,Schehr06SOS} and record statistics \cite{GodrecheMecords15,GodrecheMecords17}.  As its name suggests, a bridge random walk $X_m$ is a discrete-time process that evolves locally as in (\ref{eq:XmI}) but is constrained to return to its initial position after a fixed number of steps $n$ (see figure \ref{fig:bridge}):
\begin{align}
    X_n = X_0=0\,.\label{eq:bridgecond}
\end{align}
In the large $n$ limit, analogously to the free RW, bridge RWs with finite variance jump distributions converge towards Brownian bridges, i.e. Brownian motions that are constrained to return to their initial position after a fixed time \cite{Billingsley68}. Therefore, using known results for the expected maximum of a Brownian bridge, we anticipate that, at leading order for large $n$, ${\mathbb E}[M_n] \sim \sqrt{\pi n \sigma^2/8}$ for a bridge RW. Two natural questions then arise: how does the simple formula for the expected maximum (\ref{eq:maxfree}) generalizes to the case of bridge RWs and what are the leading terms of ${\mathbb E}[M_n]$ in the large $n$ limit? The main goal of this paper is to answer these question for bridge RWs with arbitrary jump distributions, including the case of L\'evy flights. In addition, we will show that, for jump distributions with a finite variance (with additional regularity properties discussed below, see phase I in figure \ref{fig:phaseDiagram}), the second leading order term is the same constant $\gamma$ given in (\ref{eq:gamma}) as in the free random walk, a result which does not have any simple and intuitive explanation. Interestingly, we show that this constant $\gamma$ appears in the large $n$ limit of the scaled difference between the expected area of the triangle linking the maximum and the bridge end points, and the expected absolute area under a bridge random walk (see figure \ref{fig:bridge}), namely
\begin{align}
\lim_{n \to \infty}  \frac{\frac{n}{2}\,{\mathbb E}[M_n]-\sum_{m=0}^n {\mathbb E}[|X_m|]}{\sigma\,n} = \frac{\gamma}{2} \;.\label{eq:thermo}
\end{align}
Hence the constant $\gamma$ carries the signature of the discreteness of the process, which persists even in the limit $n \to \infty$. 

\subsection{Summary of the main results}
\label{eq:intavg}
It is useful to summarize our main results. We find that the expected maximum of a bridge RW after $n$ steps is given by
\begin{align}
 {\mathbb E}[M_n] = \sum_{m=1}^n \frac{1}{m}
   \int_{0}^\infty dy\, y\, \frac{P(y,m)  P(y,n-m)}{P(0,n)}\,,\label{eq:maxbridgeI}
   \end{align}
 where $P(y,m)$ is the propagator of the free random walk (\ref{eq:PI}). This expression nicely extends the one for the expected maximum of a free random walk (\ref{eq:maxfree}). The similarity between these two formulae (\ref{eq:maxfree}) and (\ref{eq:maxbridgeI}) is further discussed in section \ref{sec:avgmaxe}. Our derivation of the expected maximum (\ref{eq:maxbridgeI}) is based on the celebrated Spitzer's formula \cite{Spitzer57} and is in agreement with a similar expression found in the mathematics literature \cite{Kac54} derived using combinatorial arguments. 
 
\begin{figure}[t]
  \begin{center}
    \includegraphics[width=0.5\textwidth]{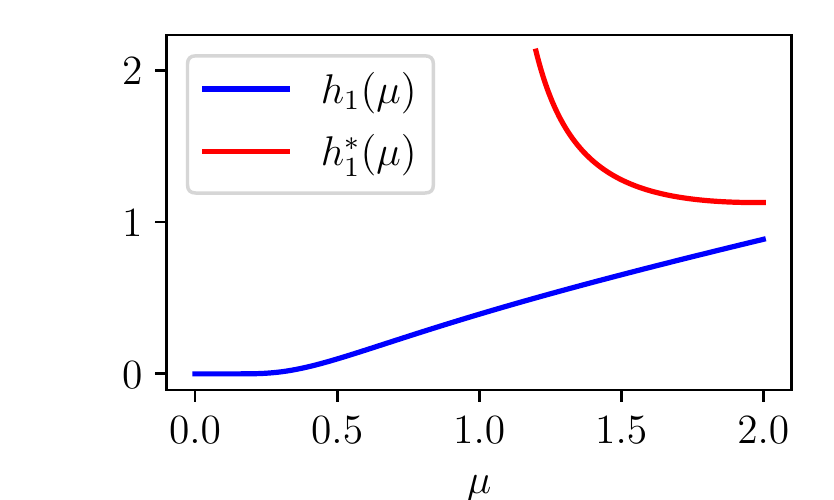}
    \caption{Amplitudes $h_1(\mu)$ in Eq. (\ref{eq:h1intro}) and $h_1^*(\mu)$ in Eq. (\ref{eq:hfree1}) of the leading order term of the large $n$ limit of the expected maximum ${\mathbb E}[M_n]$ of a bridge random walk and a free random walk, respectively as a function of the L\'evy index $0 < \mu \leq 2$. Contrary to free RWs, bridge RWs with $\mu<1$, have a well defined expected maximum due to the bridge constraint that pins the initial and final positions to the origin. This explains why the blue curve is defined for $\mu<1$ while the red one diverges upon approaching $\mu=1$.}
    \label{fig:h1}
  \end{center}
\end{figure}

 Let us now present the asymptotic behavior of the expected maximum $ {\mathbb E}[M_n]$ for a generic jump distribution $f(\eta)$.  We assume a general expansion of the Fourier transform of the jump distribution $\hat f(k)$ of the form (using similar notations to \cite{GrebenkovConvex17})
\begin{align}
  \hat f(k) \sim 1 - |ak|^\mu + b\,|k|^{\nu} + O(|k|^{2\,\mu})\,,\quad k\rightarrow 0\,,\label{eq:f}
\end{align}
where $0<\mu\leq 2$ and $\mu<\nu\leq 2\,\mu$ as well as $a>0$ and $b$ are constants. While $\mu=2$ corresponds to finite variance distributions, $\mu<2$ and $\mu<1$ corresponds to infinite variance and infinite first moment distributions respectively, with fat tails that decay like $f(\eta)\sim\eta^{-1-\mu}$ for $\eta\rightarrow\infty$. The leading term in the large $n$ limit of the expected maximum ${\mathbb E}[M_n]$ depends only on the exponent $\mu$ and the length scale $a$ and is given by
\begin{align}
   {\mathbb E}[M_n]   &\sim a\,h_1(\mu)\,n^{\frac{1}{\mu}} \,,\label{eq:Emnintro1}
\end{align}
where the amplitude $h_1(\mu)$ is given by
\begin{align}
  h_1(\mu) = \frac{\mu\pi}{8\,\Gamma\left(1+\frac{1}{\mu}\right)}\,,\label{eq:h1intro}
\end{align}
where $\Gamma(z)$ is the standard Gamma function. A plot of this function $h_1(\mu)$ is shown in figure \ref{fig:h1}. 

The second leading order term in the large $n$ limit of the expected maximum ${\mathbb E}[M_n]$ is slightly more subtle as it depends also on the exponent $\nu$ in the expansion of the jump distribution (\ref{eq:f}). We find that it displays a rich behavior depending on the two exponents $\mu$ and $\nu$ (see figure \ref{fig:phaseDiagram}):
\begin{figure}[t]
  \begin{center}
    \includegraphics[width=0.5\textwidth]{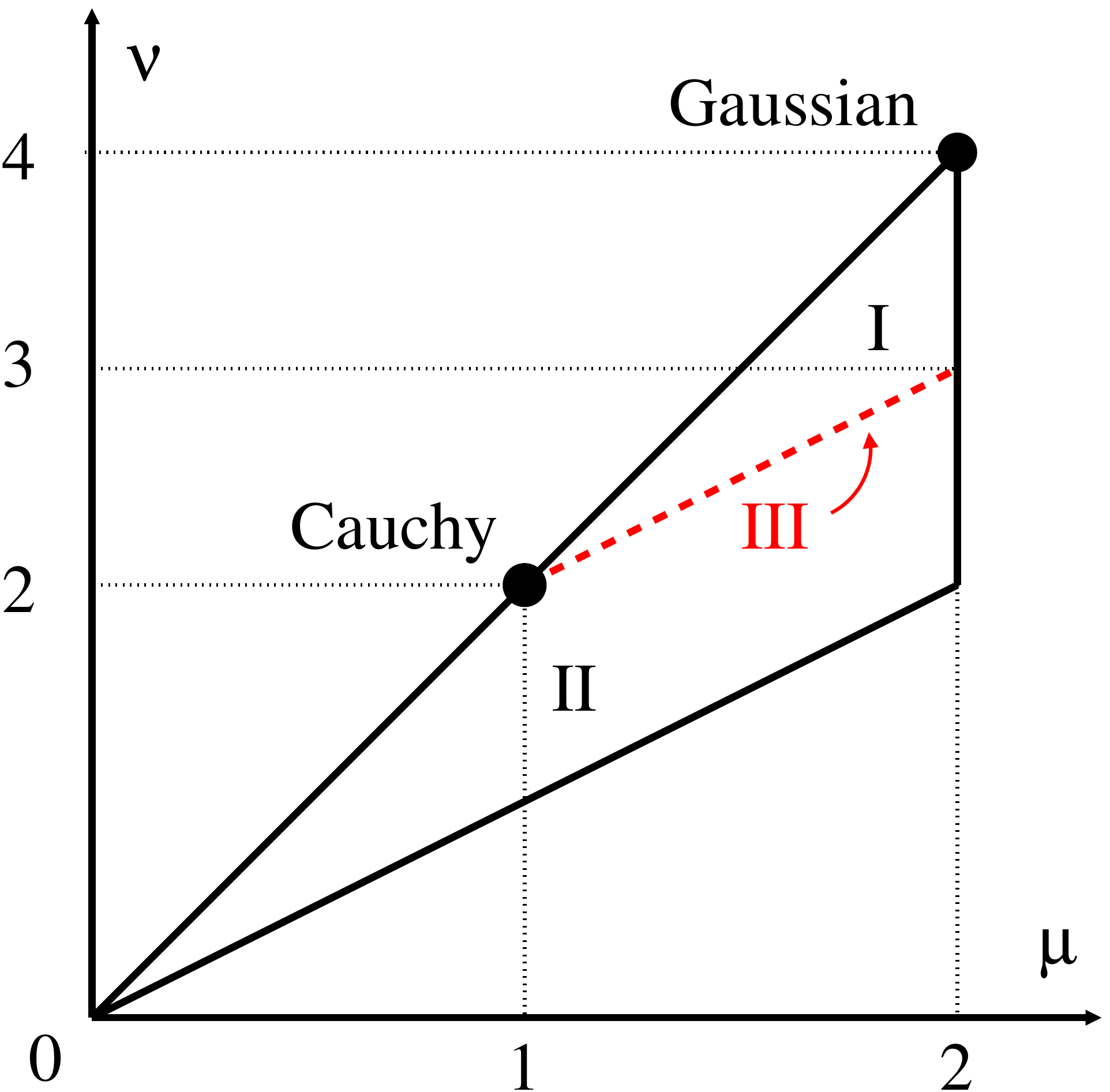}
    \caption{The parameters $\mu$ and $\nu$ are the exponents of the expansion of the Fourier transform of the jump distribution $\hat f(k)$ such that $\hat f(k) \sim 1 - a^\mu\,|k|^\mu + b\,|k|^{\nu}$ for $k\rightarrow 0$. The second leading order term in the large $n$ limit of the expected maximum ${\mathbb E}[M_n]$ depends on the phase in which the couple $(\mu,\nu)$ is located (I, II or III). The second leading order term of each phases are given in (\ref{eq:Masall}). As a reference, the couple $(\mu,\nu)$ for Gaussian and Cauchy  jump distributions are indicated at $(\mu=2,\nu=4)$ and $(\mu=1,\nu=2)$, respectively.}
    \label{fig:phaseDiagram}
  \end{center}
\end{figure}
\begin{subequations}
\begin{itemize}
  \item  Phase I ($1<\mu\leq 2$ and $\mu+1<\nu\leq 2 \mu$)
\begin{align}
 {\mathbb E}[M_n]   &\sim a\,h_1(\mu)\,n^{\frac{1}{\mu}} 
+ \frac{1}{2\pi}\int_{-\infty}^\infty  \frac{dk_1}{k_1^2} \ln\left(\frac{1-\hat f(k_1)}{a^\mu\,|k_1|^\mu}\right) \,,\quad n\rightarrow \infty\,, \label{eq:MasI}
\end{align}
\item Phase II ($0<\mu\leq 2$ and $\mu<\nu \leq 2\,\mu$ and  $\nu<\mu+1$)
\begin{align}
  {\mathbb E}[M_n]  &\sim a\,h_1(\mu)\,n^{\frac{1}{\mu}} 
  + a^{1-\nu}\,b \,h_2(\mu,\nu)\,n^{\frac{1+\mu-\nu}{\mu}}\,,\quad n\rightarrow \infty\,,\label{eq:MasII}
\end{align}
\item Phase III ($1\leq\mu\leq 2$ and $\nu=\mu+1$)
\begin{align}
 {\mathbb E}[M_n] &\sim a\,h_1(\mu)\,n^{\frac{1}{\mu}}  -\frac{b}{\pi\,\mu\,a^{\mu}}\,\ln(n)\,,\quad n\rightarrow \infty\,,\label{eq:MasIII}
    \end{align}
\end{itemize}
\label{eq:Masall}
\end{subequations}
where the amplitude $h_1(\mu)$ is given in (\ref{eq:h1intro}) and $h_2(\mu,\nu)$ is given by 
\begin{align}
  h_2(\mu,\nu) = \frac{-1}{2\,\Gamma\left(1+\frac{1}{\mu}\right)}\left(\frac{\pi \, \Gamma \left(\frac{\nu +1}{\mu }\right)}{4\, \Gamma
   \left(1+\frac{1}{\mu }\right)} + \frac{  (\nu-\mu  )  \csc
   \left(\frac{\pi  \nu }{\mu }\right) }{\mu  
   \Gamma \left(-\frac{\nu }{\mu }\right)}\int_0^\infty dv \frac{\left(v^{\nu }-1\right)\left(v^{\mu -\nu
   }-1\right)}{\left(v^2-1\right) \left(v^{\mu }-1\right)}\right)\,.\label{eq:h2s}
\end{align}
Note that the amplitude $h_2(\mu,\nu)$ diverges as $\nu$ approaches $\mu+1$ as expected as it corresponds to approaching the red line in figure \ref{fig:phaseDiagram}. The amplitude $h_2(\mu,\nu)$ can be exactly evaluated for $\mu=1$ and $\mu=2$, leading to
\begin{subequations}
\begin{align}
  h_2(\mu=1,\nu) &=  \frac{ \pi ^2-2 (\nu -1) [\pi  (2 \nu -1)
   \cot (\pi  \nu )+\pi  \csc (\pi  \nu )-2]}{8\sin (\pi  \nu ) \Gamma
   (-\nu )}\, ,\label{eq:h21}\\
h_2(\mu=2,\nu) &=  \frac{\sqrt{\pi } (\nu -2) (\nu -1) \sec
   \left(\frac{\pi  \nu }{2}\right)}{4\,\Gamma \left(-\frac{\nu
   }{2}\right)}-\frac{1}{2} \Gamma \left(\frac{\nu +1}{2}\right)\,.\label{eq:h22}
\end{align}    
\end{subequations}
As discussed below in Section \ref{sec:asMn}, we have verified numerically our results on several jump distributions using a numerical method that was recently proposed in \cite{DebruyneBridge21}. 
 
It is interesting to compare the asymptotic large $n$ expansion of the expected maximum of a bridge RW (\ref{eq:Masall}) with the one of a free RW obtained previously in \cite{ComtetPrecise05,GrebenkovConvex17}. While similar phases and exponents were obtained in both cases, the results for the bridge RW differ from the free RW in two ways: (i) ${\mathbb E}[M_n]$ for the bridge RW is finite even for L\'evy flights with L\'evy exponent $0<\mu \leq 1$ (lower left region in phase II) while it is infinite for a free RW: this 
is due to the bridge constraint that pins the initial and final positions to the origin, (ii) the amplitudes $h_1(\mu)$ and $h_2(\mu,\nu)$ are different from the ones obtained  for the free random walks $h^*_1(\mu)$ and $h^*_2(\mu,\nu)$ given by \cite{ComtetPrecise05,GrebenkovConvex17}
\begin{subequations}
\begin{align}
  h^*_1(\mu) &= \frac{\mu}{\pi}\,\Gamma\left(1-\frac{1}{\mu}\right)\,,\label{eq:hfree1}\\
  h^*_2(\mu,\nu) &= \frac{1}{\pi\left(\nu-\mu-1\right)}\,\Gamma\left(\frac{\nu-1}{\mu}\right)\,.\label{eq:hfree2}
\end{align}  
\label{eq:hfree}
\end{subequations}
Comparing the amplitude $h_1(\mu)$ and $h_1^*(\mu)$ (see figure \ref{fig:h1}), we see that the amplitude of the free RW (for $\mu > 1$) is larger than the one for the bridge constraint, which is expected as the bridge RW cannot go as far as the free RW due to its constraint to return to the origin.

In the specific case of a jump distribution with finite variance $\sigma$ in phase I, the asymptotic limit of the expected maximum of the bridge random walk is given by
\begin{align}
  \frac{{\mathbb E}[M_n]}{\sigma}   &\sim \sqrt{\frac{\pi\,n}{8}} + \gamma\, \quad, \quad n \to \infty\,,\label{eq:maxga}
\end{align}
where, remarkably, $\gamma$ is the same constant correction (\ref{eq:gamma}) as in the expected maximum of the free RW obtained in \cite{ComtetPrecise05} discussed in the introduction. It turns out that this finite size correction appears in the large $n$ limit of the scaled difference between the expected area of the triangle linking the maximum and the bridge end points and the expected absolute area under a bridge random walk (see figure \ref{fig:bridge}). In section \ref{sec:area}, we show that in the large $n$ limit, the leading first order terms of these two quantities exactly coincide, such that the leading order term of their difference becomes
\begin{align}
 \lim_{n \to \infty} \frac{\frac{n}{2}\,{\mathbb E}[M_n]-\sum_{m=0}^n {\mathbb E}[|X_m|]}{\sigma\,n}= \frac{\gamma}{2} \,.\label{eq:thermoi}
\end{align}

The rest of the paper is organized as follows.  In Section \ref{sec:avgmax}, we study the expected maximum of a bridge RW with an arbitrary symmetric jump distribution and discuss its large $n$ asymptotic limit depending on the parameters $\mu$ and $\nu$ of the jump distribution in (\ref{eq:f}). In this section, we also present a comparison of these analytical results with numerical simulations. We close this section by establishing the connection with the large $n$ limit of the expected absolute area under the bridge, leading to the relation~(\ref{eq:thermo}). In Section \ref{sec:app}, we discuss applications of our results to the convex hull of polymer chains \cite{ComtetPrecise05}, the convex hull of the trajectory of $2d$ run-and-tumble particles \cite{HartmannConvex20,Majumdar21convex} and to the survival probability in 
a bridge version of the ``lamb-lion'' problem \cite{Majumdar19Smol,Franke12,Krapivsky96Lamb,Redner99Lamb,Redner01}. Finally, Section \ref{sec:ccl} contains our conclusions and perspectives. Some detailed calculations are presented in \ref{app:h1} and \ref{app:2max}.

\section{Expected maximum of a bridge random walk}
\label{sec:avgmax}
\subsection{Exact result for finite $n$}
\label{sec:avgmaxe}
In this section, we derive the exact expression (\ref{eq:maxbridgeI}) for the expected maximum of a symmetric bridge random walk presented in the introduction. The derivation is based on Spitzer's formula \cite{Spitzer57} which gives us the generating function of the double Laplace-Fourier transform of the joint distribution $P_{\text{joint}}(M_n,x_n,n)$ of the maximum $M_n$ and the final position $x_n$ of a free random walk after $n$ steps. Assuming that the jump distribution $f(\eta)$ is symmetric, Spitzer's formula reads \cite{Spitzer57,WendelS58}
\begin{align}
  \sum_{n=0}^\infty z^n\, {\mathbb E}[\rme^{-\alpha M_n+ik x_n}]&= \exp\left[\sum_{n=1}^\infty \frac{z^n}{n}
   \int_{0}^\infty dy\, P(y,n) \left(\rme^{(ik- \alpha) y}+\rme^{-i k y}\right)\right]\,,\label{eq:spitzer}
\end{align}
where ${\mathbb E}[\rme^{-\alpha M_n+ik x_n}]$ is the double Laplace-Fourier transform of the joint distribution $P_\text{joint}(M_n,x_n,n)$, and $P(x_n,n)=\int_0^\infty dM_n P_\text{joint}(M_n,x_n,n)$ is the forward propagator of the free random walk defined in (\ref{eq:PI}). Taking a derivative with respect to $\alpha$ and setting $\alpha=0$ in (\ref{eq:spitzer}) gives
\begin{align}
 &\sum_{n=0}^\infty z^n \,{\mathbb E}[M_n\,\rme^{ik x_n}] \nonumber \\
 &= \sum_{m=1}^\infty \frac{z^m}{m}
   \int_{0}^\infty dy\, y\, P(y,m) \rme^{ik y}\exp\left[\sum_{n=1}^\infty \frac{z^n}{n}
   \int_{0}^\infty dy \,P(y,n) \left(\rme^{ik y}+\rme^{-i k y}\right)\right]\,.\label{eq:spitzerder}
\end{align}
In the argument of the exponential in the right hand side in (\ref{eq:spitzerder}), we recognize the Fourier transform of the forward propagator $\int_{-\infty}^{\infty}dy \,P(y,n)\,e^{iky}$. 
Inserting the expression of the Fourier transform of the propagator (\ref{eq:PI}) in the argument of the exponential in (\ref{eq:spitzerder}) and performing the sum over $n$ gives
\begin{align}
 \exp\left[\sum_{n=1}^\infty \frac{z^n}{n}
   \int_{0}^\infty dy\, P(y,n) \left(\rme^{ik y}+\rme^{-i k y}\right)\right]&= \exp\left[\sum_{n=1}^\infty \frac{z^n}{n}
   \hat f(k)^n\right]\nonumber\,,\\ &=   \exp\left[-\ln(1-z\,\hat f(k))\right]\,,\nonumber\\
   &= \frac{1}{1-z\,\hat f(k)} =\sum_{l=0}^\infty  \hat f(k)^l z^l\,,\label{eq:sumexp}
\end{align}
where $\hat f(k)$ is the Fourier transform of the jump distribution. Plugging back the relation (\ref{eq:sumexp}) into the generating function (\ref{eq:spitzerder}) yields
\begin{align}
  \sum_{n=0}^\infty z^n \,{\mathbb E}[M_n\,\rme^{ik x_n}] = \sum_{m=1}^\infty \frac{z^m}{m}
   \int_{0}^\infty dy\, y\, P(y,m) \rme^{ik y}\sum_{l=0}^\infty  \hat f(k)^l z^l\,.\label{eq:powz}
\end{align}
Identifying the power of $z$ in the left and the right hand side of (\ref{eq:powz}) gives
\begin{align}
 {\mathbb E}[M_n\,\rme^{ik x_n}]= \sum_{m=1}^n \frac{1}{m}
   \int_{0}^\infty dy\, y\, P(y,m)\, \rme^{ik y}\, \hat f(k)^{n-m}\,.\label{eq:powzid}
\end{align}
Applying an inverse Fourier transform on both sides, we obtain
\begin{align}
\int_{0}^{\infty} dM_n \,M_n\,\,P(M_n,x_n,n)&= \sum_{m=1}^n \frac{1}{m}
   \int_{0}^\infty\,dy\, y\, P(y,m) \, P(x_n-y,n-m)\,.\label{eq:M}
\end{align}
Upon setting the final position to be at the origin $x_n=0$ to satisfy the bridge condition (\ref{eq:bridgecond}) and denoting ${\mathbb E}[M_n]$ the expected maximum for the bridge after $n$ jumps, we finally obtain the expression (\ref{eq:maxbridgeI}) presented in the introduction, i.e., 
\begin{align}
   {\mathbb E}[M_n]
  &= \frac{\int_{0}^{\infty} dM_n \,M_n\,\,P(M_n,0,n) }{\int_{0}^{\infty} dM_n \,\,P(M_n,0,n) }=
    \frac{\sum_{m=1}^n \frac{1}{m}\int_{0}^\infty dy\, y\,P(y,m)  P(y,n-m)}{P(0,n)}\,,\label{eq:Mn}
\end{align}
where we used the symmetry $P(y,n)=P(-y,n)$ as the random walk has a symmetric jump distribution. The expected maximum (\ref{eq:Mn}) has the same functional form as the expected maximum for a free random walk (\ref{eq:maxfree}) as it is given by a weighted sum over $m$ of the expected absolute position of the random walk. Indeed, the bridge propagator $P_\text{bridge}(X,m\,|\,n)$, i.e., the probability density that the random walk is located at $X$ at step $m$ given that it must return to the origin after $n$ steps, is given by the normalized product of two free propagators (see figure \ref{fig:bridge}):
\begin{align}
  P_\text{bridge}(y,m\,|\,n) = \frac{P(y,m)  P(y,n-m)}{P(0,n)}\,,\label{eq:Pbridge}
\end{align}
where the first propagator accounts for the left part of the bridge random walk of length $m$ from the origin to $y$ and the second one accounts for the right part of the bridge random walk of length $n-m$ from $y$ back to the origin. The normalization factor $P(0,n)$ in the denominator accounts for  all the bridge trajectories of length $n$. Therefore, the expression (\ref{eq:Mn}) can be written as ${\mathbb E}[M_n]=\sum_{m=1}^n \frac{1}{m}\int_{0}^\infty dy\, y\,P_\text{bridge}(y,m\,|\,n)$ and nicely extends the equation (\ref{eq:maxfree}) for the expected maximum of free RWs to the case of bridge RWs.

\subsection{Asymptotic results for large $n$}
\label{sec:asMn}
We study the large $n$ limit of the expected maximum of a symmetric bridge random walk (\ref{eq:Mn}) up to second leading order. We only present here, in the main text, the derivation of the first leading order term and provide the detailed calculations of the second leading order in \ref{app:2max}. We end this section by verifying our results numerically on various jump distributions.
\subsubsection{Leading order}
To study the large $n$ limit of the expected maximum, we analyze the numerator and the denominator in (\ref{eq:Mn}) separately.  To analyze the numerator, it is useful to introduce its generating function 
\begin{align}
g(z) \equiv \sum_{n=0}^\infty z^n \sum_{m=1}^n \frac{1}{m}\int_{0}^\infty dy\, y\,P(y,m)  P(y,n-m)\,.\label{eq:g}
\end{align}
Replacing the propagator $P(y,n)$ by its Fourier transform (\ref{eq:PI}), we find
\begin{align}
  g(z) &= \sum_{n=0}^\infty z^n\sum_{m=1}^n \frac{1}{m} \int_{-\infty}^\infty \frac{dk_1}{2\pi} \int_{-\infty}^\infty \frac{dk_2}{2\pi} \hat f(k_1)^m \, \hat f(k_2)^{n-m}\,\int_0^\infty dy\, y\, \rme^{-i(k_1+k_2)\,y} \,.\label{eq:gzMn}
\end{align}
The integral over $y$ can now be performed and yields
\begin{align}
    g(z) &= -\sum_{n=0}^\infty z^n \sum_{m=1}^n \frac{1}{m} \int_{-\infty}^\infty \frac{dk_1}{2\pi} \int_{-\infty}^\infty \frac{dk_2}{2\pi} \hat f(k_1)^m \, \hat f(k_2)^{n-m}\,\frac{1}{(k_1+k_2-i\epsilon)^2}\,,\label{eq:befored}
\end{align}
where we regularized the integral with a regularization parameter $\epsilon$ that must be taken to $\epsilon\rightarrow 0^+$ after performing the integration. The sum over $n$ and $m$ can now be performed and we obtain
\begin{align}
 g(z)&=\frac{1}{4\pi^2}\int_{-\infty}^\infty \int_{-\infty}^\infty dk_1 dk_2 \frac{\ln(1-z\hat f(k_1))}{(k_1+k_2-i\epsilon)^2} \frac{1}{1-z\hat f(k_2)}\,.\label{eq:g2sc}
\end{align}
We now see that in the limit $z\rightarrow 1$, the integrals in the generating function (\ref{eq:g2sc}) are dominated for $k_1\rightarrow 0$ and $k_2\rightarrow 0$. Indeed, the Fourier transform of the jump distribution behaves like $\hat f(k)\rightarrow 1-a^\mu |k|^\mu$ for $k\rightarrow 0$, which makes the integral diverge in the limit $z\rightarrow 1$. We therefore replace $\hat f(k)$ by its small $k$ expansion (\ref{eq:f}) and rescale $k_1$ and $k_2$ by $(1-z)^{1/\mu}/a$ to find that the generating function behaves, to leading order,~as
\begin{align}
  g(z)&\sim  \frac{1}{4\pi^2(1-z)}\int_{-\infty}^\infty \int_{-\infty}^\infty dk_1 dk_2 \frac{\ln(1+|k_1|^\mu)}{(k_1+k_2-i\epsilon)^2(1+|k_2|^\mu)}\,\,,\quad z\rightarrow 1\,.\label{eq:gm}
\end{align}
We now invert the generating function (\ref{eq:g}) by using the Tauberian theorem
\begin{align}
  \sum_{n=0}^\infty z^n a_n &\sim\frac{1}{(1-z)^\alpha}\,,\quad z\rightarrow 1\,,\quad \iff \quad a_n \sim \frac{n^{\alpha-1}}{\Gamma(\alpha)}\,,\quad n\rightarrow\infty\,,\label{eq:taub}
\end{align}
to find that the leading order of the numerator in (\ref{eq:Mn}) behaves, as $n \to \infty$, as 
\begin{align}
  \sum_{m=1}^n \frac{1}{m}\int_{0}^\infty dy\, y\,P(y,m)  P(y,n-m) &\sim\frac{1}{4\pi^2}\int_{-\infty}^\infty \int_{-\infty}^\infty dk_1 dk_2 \frac{\ln(1+|k_1|^\mu)}{(k_1+k_2-i\epsilon)^2(1+|k_2|^\mu)}\,.\label{eq:gma}
\end{align}
The denominator in (\ref{eq:Mn}) can be analyzed by replacing the propagator $P(y,n)$ by its Fourier transform (\ref{eq:PI}) to find that
\begin{align}
P(y=0,n)=\int_{-\infty}^\infty \frac{dk}{2\pi} \hat f(k)^n &=  \int_{-\infty}^\infty \frac{dk}{2\pi} \rme^{ n \log(\hat f(k))} \,.\label{eq:denm}
\end{align}
For large $n$, the integral is dominated for $k\rightarrow 0$ since $\hat f(k)\rightarrow 1-a^\mu |k|^\mu$ when $k\rightarrow 0$. We can therefore replace $\hat f(k)$ by its small $k$ expansion (\ref{eq:f}) which gives that the denominator in (\ref{eq:Mn}) behaves as
\begin{align}
 P(y=0,n)&\sim \int_{-\infty}^\infty \frac{dk}{2\pi} \rme^{ -n\left(a^\mu |k|^{\mu}\right) }\sim \frac{1}{\pi \,a }\,\Gamma \left(1+\frac{1}{\mu }\right)  n^{-\frac{1}{\mu} }\,,\quad n\rightarrow\infty\,.\label{eq:denlm}
   \end{align}
Combining the asymptotic limits of the numerator (\ref{eq:gma}) and the denominator (\ref{eq:denlm}), we find that the leading order of the expected maximum of a bridge random walk (\ref{eq:Mn}) is
\begin{align}
 {\mathbb E}[M_n]  &\sim h_1(\mu)\,a\,n^{\frac{1}{\mu}}  \,,\quad n\rightarrow \infty\,, \label{eq:asMnm}
\end{align}
where the amplitude $h_1(\mu)$ is given by
\begin{align}
  h_1(\mu) = \frac{1}{4\,\pi\,\Gamma\left(1+\frac{1}{\mu}\right)}\int_{-\infty}^\infty \int_{-\infty}^\infty dk_1 dk_2 \frac{\ln(1+|k_1|^\mu)}{(k_1+k_2-i\epsilon)^2(1+|k_2|^\mu)}\,.\label{eq:h1}
\end{align} Quite remarkably, it is possible to perform this double integral (see \ref{app:h1}) and one obtains finally the expression given in (\ref{eq:h1intro}) in the introduction.

For the second leading term, the detailed calculations can be found in \ref{app:2max}. One needs to consider the second leading term in the expansion of the Fourier transform of the jump distribution $\hat f(k) \sim 1 - a^\mu\,|k|^\mu + b\,|k|^{\nu}$ when $k\rightarrow 0$. For $1<\mu\leq 2$ and $\mu+1<\nu\leq 2 \mu$ (phase I in figure \ref{fig:phaseDiagram}), the derivation is given in \ref{app:case1} and the result is given in (\ref{eq:MasI}). For $0<\mu\leq 2$, $\mu<\nu \leq 2\,\mu$, and  $\nu<\mu+1$ (phase II in figure \ref{fig:phaseDiagram}), the derivation is given in \ref{app:case2} and the result is given in (\ref{eq:MasII}). Finally, for $1\leq\mu\leq 2$ and $\nu=\mu+1$ (phase III in figure \ref{fig:phaseDiagram}), the derivation is given in \ref{app:case3} and the result is given in (\ref{eq:MasIII}).
\subsubsection{Numerical results}
We verify our results for the asymptotic maximum (\ref{eq:Masall}) by generating bridge RWs with various jump distributions that belong to different regions in the phase diagram in figure \ref{fig:phaseDiagram}. We employ  a recently devised method to generate bridge RWs \cite{DebruyneBridge21}. Below, we report the numerical estimation of the expected maximum ${\mathbb E}[M_n]_{\text{num}}$ and the ratio of the difference between ${\mathbb E}[M_n]_{\text{num}}$ and the first leading order term over the second leading order term, which should tend to $1$ as $n\rightarrow\infty$.

\paragraph{Phase I ($1<\mu\leq 2$ and $\mu+1<\nu\leq 2 \mu$):}
To numerically verify the asymptotic results for the expected maximum in phase I, we choose a centered Gaussian jump distribution with variance $\sigma$
\begin{align}
    f(\eta) = \frac{1}{\sqrt{2\pi\,\sigma^2}}\,\rme^{-\frac{\eta^2}{2\sigma^2}}\label{eq:gaussDist}\,,
\end{align}
 which belongs to phase I in the diagram in figure \ref{fig:phaseDiagram} as its Fourier transform is
\begin{align}
  \hat f(k) = \rme^{-\frac{\sigma^2\,k^2}{2}} \sim 1  -\frac{\sigma^2}{2}\,k^2 + \frac{\sigma^4}{8}\,k^4\,,\quad k \rightarrow 0\,.\label{eq:f0G}
\end{align}
We read off the parameters in the general expansion (\ref{eq:f}) from (\ref{eq:f0G}) to be
\begin{align}
  \mu =2\,,\quad \nu =4\,,\quad a = \frac{\sigma}{\sqrt{2}}\,,\quad b = \frac{\sigma^4}{8}\,,\label{eq:parG}
\end{align}
which corresponds to the upper right corner in the phase diagram in figure \ref{fig:phaseDiagram}. We generate Gaussian bridges using the recently proposed algorithm in \cite{DebruyneBridge21} and report the ratio $r(n)$ of the difference between ${\mathbb E}[M_n]_{\text{num}}$ and the first leading order term over the second leading order term
\begin{align}
  r(n) = \frac{ {\mathbb E}[M_n]_{\text{num}} - a\,h_1(\mu)\,n^{\frac{1}{\mu}} }{\frac{1}{2\pi}\int_{-\infty}^\infty  \frac{dk_1}{k_1^2} \ln\left(\frac{1-\hat f(k_1)}{a^\mu\,|k_1|^\mu}\right)}\,.\label{eq:rI}
\end{align}
 We find that the asymptotic results for the expected maximum (\ref{eq:MasI}) are in excellent agreement with numerical data (see Table \ref{table:bridgeg}).

\begin{table}[t]
\centering
\begin{tabular}{ |l|l|l|  }
\hline
$n$ & ${\mathbb E}[M_n]_{\text{num}}$ & $r(n)$\\
\hline\hline
$10$ & $1.3944\pm 3\times 10^{-4}$   &$1.0082\pm 5\times 10^{-4}$  \\
$100$ & $5.683\pm 1\times 10^{-3}$ & $1.001\pm 2\times 10^{-3}$\\
$1000$ & $19.232\pm 3\times 10^{-3}$ & $1.003\pm 5\times 10^{-3}$ \\
\hline
\end{tabular}
\caption{Expected maximum of a Gaussian bridge random walk of $n$ jumps ($\sigma=1$, $10^7$ runs). The quantity $r(n)$ is the ratio between the numerical second leading term and the theoretical one (\ref{eq:rI}). }
\label{table:bridgeg}
\end{table}

\paragraph{Phase II ($0<\mu\leq 2$ and $\mu<\nu \leq 2\,\mu$ and  $\nu<\mu+1$):}
To numerically verify the asymptotic results for the expected maximum in phase II (\ref{eq:MasII}), we choose the Student's t-distribution of parameter $5/2$, namely
\begin{align}
   f(\eta) = \sqrt{\frac{2}{5 \pi }}\frac{ \Gamma \left(\frac{7}{4}\right)}{
    \Gamma \left(\frac{5}{4}\right)}\,\frac{1}{\left(\frac{2 \eta ^2}{5}+1\right)^{7/4}}\,,\label{eq:studD}
\end{align} which belongs to phase II in the diagram in figure \ref{fig:phaseDiagram} as its Fourier transform is
\begin{align}
  \hat f(k) &= \Gamma \left(\frac{3}{4}\right)\,\frac{10^{5/8}}{\pi }\, | k| ^{5/4}\,
   K_{\frac{5}{4}}\left(\sqrt{\frac{5}{2}}\, | k| \right)\nonumber\\
   &\sim 1-\frac{5 }{2}\,k^2-\frac{1}{\pi
   }\,\sqrt[4]{\frac{5}{2}} \Gamma
   \left(-\frac{1}{4}\right) \Gamma \left(\frac{3}{4}\right)\, |k|^{5/2}\,,\quad k \rightarrow 0\,.\label{eq:f0stud}
\end{align}
We read off the parameters in the general expansion (\ref{eq:f}) from (\ref{eq:f0stud}) to be
\begin{align}
  \mu =2\,,\quad \nu =\frac{5}{2}\,,\quad a = \sqrt{\frac{5}{2}}\,,\quad b = -\frac{1}{\pi
   }\,\sqrt[4]{\frac{5}{2}} \Gamma
   \left(-\frac{1}{4}\right) \Gamma \left(\frac{3}{4}\right)\,.\label{eq:parstud}
\end{align}
 We generated Student's t-distribution bridge RWs as in \cite{DebruyneBridge21} and recorded their maximum. We then computed the numerical average maximum ${\mathbb E}[M_n]_{\text{num}}$  and report the ratio $r(n)$ of the difference between ${\mathbb E}[M_n]_{\text{num}}$ and the first leading order term over the second leading order term
\begin{align}
  r(n) = \frac{ {\mathbb E}[M_n] - a\,h_1(\mu)\,n^{\frac{1}{\mu}} }{a^{1-\nu}\,b \,h_2(\mu,\nu)\,n^{\frac{1+\mu-\nu}{\mu}}}\,.\label{eq:rII}
\end{align}
 We find that the asymptotic results for the expected maximum (\ref{eq:MasII}) is in good agreement with numerical data (see Table \ref{table:bridges}).
\begin{table}[t]
\centering
\begin{tabular}{ |l|l|l|  }
\hline
$n$ & ${\mathbb E}[M_n]$ & $r(n)$\\
\hline\hline
$10$ & $1.9388\pm 5\times 10^{-4}$   &$1.754\pm 2\times 10^{-4}$  \\
$100$ & $9.858\pm 1\times 10^{-3}$ & $1.2893\pm 6\times 10^{-4}$\\
$1000$ & $37.477\pm 7\times 10^{-3}$ & $1.193 \pm 1\times 10^{-3}$ \\
\hline
\end{tabular}
\caption{Expected maximum of a bridge random walks of $n$ jumps ($10^7$ runs) with a Student's t-jump distribution (\ref{eq:studD}). The quantity $r(n)$ is the ratio between the numerical second leading term and the theoretical one (\ref{eq:rII}).}
\label{table:bridges}
\end{table}

\paragraph{Phase III ($1\leq\mu\leq 2$ and $\nu=\mu+1$):} 
To numerically verify the asymptotic results for the expected maximum in phase III (\ref{eq:MasIII}), we choose a Cauchy distribution of scale $\ell$, namely
\begin{align}
  f(\eta) = \frac{1}{\ell\,\pi}\,\frac{1}{1+\left(\frac{\eta}{\ell}\right)^2}\,,\label{eq:cauchyDist}
\end{align}
 which belongs to phase III as its Fourier transform is
\begin{align}
  \hat f(k) = \rme^{-\ell |k|}\sim 1-\ell|k| + \frac{\ell^2}{2}|k|^2\,,\quad k \rightarrow 0\,.\label{eq:f0C}
\end{align}
We read off the parameters in the general expansion (\ref{eq:f}) from (\ref{eq:f0C}) to be
\begin{align}
  \mu =1\,,\quad \nu =2\,,\quad a =\ell\,,\quad b = \frac{\ell^2}{2}\,,\label{eq:parC}
\end{align}
which corresponds to the middle point of the main diagonal in the phase diagram in figure \ref{fig:phaseDiagram}. We generated Cauchy bridge RWs as in \cite{DebruyneBridge21} and recorded their maximum. We then computed the numerical expected maximum ${\mathbb E}[M_n]_{\text{num}}$ and report the ratio $r(n)$ of the difference between ${\mathbb E}[M_n]_{\text{num}}$ and the first leading order term over the second leading order term
\begin{align}
  r(n) = \frac{{\mathbb E}[M_n] - a\,h_1(\mu)\,n^{\frac{1}{\mu}} }{-\frac{b}{\pi\,\mu\,a^{\mu}}\,\ln(n)}\,.\label{eq:rIII}
\end{align}
We find that the asymptotic results for the expected maximum (\ref{eq:MasIII}) is in good agreement with numerical data (see Table \ref{table:bridgec}). The convergence is slightly slower than the other two cases, probably due to the second order logarithmic correction.

\begin{table}[t]
\centering
\begin{tabular}{ |l|l|l|  }
\hline
$n$ & ${\mathbb E}[M_n]$ & $r(n)$\\
\hline\hline
$10$ & $3.248\pm 1\times 10^{-3}$   &$1.852\pm 3\times 10^{-3}$  \\
$100$ & $38.24\pm 1\times 10^{-2}$ & $1.41\pm 1\times 10^{-2}$\\
$1000$ & $391.27\pm 6\times 10^{-2} $ & $1.30\pm 5\times 10^{-2}$ \\
$5000$ & $1961.9\pm 2\times 10^{-1} $ & $1.1\pm 2\times 10^{-1}$ \\
\hline
\end{tabular}
\caption{Expected maximum of a Cauchy bridge random walk of $n$ jumps ($10^7$ runs). The quantity $r(n)$ is the ratio between the numerical second leading term and the theoretical one (\ref{eq:rIII}).}
\label{table:bridgec}
\end{table}

In addition, we also generated a Student's t-distribution of parameter $3$, namely
\begin{align}
  f(\eta) = \frac{6 \sqrt{3}}{\pi  \left(\eta^2+3\right)^2}\,,\label{eq:stud2D}
\end{align} 
 which also belongs to phase III as its Fourier transform is
\begin{align}
  \hat f(k) = \frac{\rme^{-\sqrt{3} | k| }
   \left(3 | k|
   +\sqrt{3}\right)}{\sqrt{3}}\sim 1-\frac{3 }{2}\,k^2-\sqrt{3}\, |k|^{3}\,,\quad k \rightarrow 0\,.\label{eq:fstud2}
\end{align}
We read off the parameters in the general expansion (\ref{eq:f}) from (\ref{eq:fstud2}) to be
\begin{align}
  \mu =2\,,\quad \nu =3\,,\quad a =\sqrt{\frac{3}{2}}\,,\quad b = \sqrt{3}\,.\label{eq:parstud2}
\end{align}
Here also, we find that the asymptotic results for the expected maximum (\ref{eq:MasIII}) are in good agreement with our numerical data (see Table \ref{table:bridgesm}) with a slightly slower convergence due probably again to the second order logarithmic correction.
\begin{table}[t]
\centering
\begin{tabular}{ |l|l|l|  }
\hline
$n$ & ${\mathbb E}[M_n]$ & $r(n)$\\
\hline\hline
$10$ & $1.8355\pm 4\times 10^{-4}$   &$3.774\pm 1\times 10^{-3} $  \\
$100$ & $8.794\pm 1\times 10^{-3}$ & $2.434\pm 2\times 10^{-3}$\\
$1000$ & $32.005\pm 6\times 10^{-3}$ & $1.826\pm 5\times 10^{-3}$ \\
$5000$ & $74.91 \pm 1\times 10^{-2}$ & $1.174\pm 9\times 10^{-3} $ \\
\hline
\end{tabular}
\caption{Expected maximum of a bridge random walks of $n$ jumps ($10^7$ runs) with a Student's t-jump distribution (\ref{eq:stud2D}). The quantity $r(n)$ is the ratio between the numerical second leading term and the theoretical one (\ref{eq:rIII}).}
\label{table:bridgesm}
\end{table}

\subsection{Large $n$ limit of the expected absolute area under a bridge}
\label{sec:area}
In the previous section we have shown that the expected maximum of a bridge RW with a jump distribution with a finite variance $\sigma$ in phase I grows like
\begin{align}
  \frac{{\mathbb E}[M_n]}{\sigma}   &\sim \sqrt{\frac{\pi\,n}{8}} + \gamma\,.\label{eq:maxga2}
\end{align}
In this section, we show how the leading finite size correction $\gamma$ appears in the large $n$ limit of the scaled difference between the expected area of the triangle linking the maximum and the bridge end points (see figure \ref{fig:bridge}), and the expected absolute area under a bridge (\ref{eq:thermo}). 

The expected absolute area under a bridge random walk after $n$ jumps reads
\begin{align}
  \sum_{m=0}^n {\mathbb E}[|X_m|] = \sum_{m=0}^n \int_{-\infty}^\infty dX\, |X|\, P_\text{bridge}(X,m\,|\,n) \;, \,\label{eq:aread}
\end{align}
where the bridge propagator $P_\text{bridge}(X,m\,|\,n)$ is given in (\ref{eq:Pbridge}). Inserting its expression (\ref{eq:Pbridge}) in the expected absolute area (\ref{eq:aread}) and using the fact that the integrand is symmetric under the change $X \to -X$ to restrict the integral over the positive half line, we find
\begin{align}
  \sum_{m=0}^n {\mathbb E}[|X_m|] = 2\,\frac{\sum_{m=0}^n \int_{0}^\infty dX\, X P(X,m)  P(X,n-m)}{P(0,n)}\,.\label{eq:areap}
\end{align}
This expression is very similar to the one found for the expected maximum in (\ref{eq:Mn}), except for the $1/m$ factor that appears there. The derivation of the asymptotic limit of the expected area carries analogously to the one of the expected maximum. We define the generating function of the numerator in (\ref{eq:areap}) as
\begin{align}
  g_a(z) \equiv \sum_{n=0}^\infty z^n \sum_{m=0}^n \int_{0}^\infty dX\, X\,P(X,m)  P(X,n-m)\,.\label{eq:ga}
\end{align}
Replacing the free propagator by its Fourier transform (\ref{eq:PI}) and performing the sums gives
\begin{align}
   g_a(z) = - \frac{1}{4\pi^2}\int_{-\infty}^\infty \int_{-\infty}^\infty dk_1 dk_2 \,\frac{1}{(k_1+k_2-i\epsilon)^2} \frac{1}{1-z\hat f(k_1)}\frac{1}{1-z\hat f(k_2)}\,,\label{eq:gazf}
\end{align}
where $\epsilon$ is a regularization parameter.
In the limit $z \rightarrow 1$, the integrals are dominated for $k_1\rightarrow 0$ and $k_2\rightarrow 0$. We therefore replace $\hat f(k)$ by its small $k$ expansion, which for a jump distribution with a finite variance $\sigma$ is
\begin{align}
  \hat f(k) \sim 1 - \frac{\sigma^2}{2}\,k^2 + O(k^4)\,,\label{eq:fks}
\end{align}
where we have also assumed a finite fourth moment. Inserting (\ref{eq:fks}) in (\ref{eq:gazf}) and rescaling $k_1$ and $k_2$ by $\sqrt{2}(1-z)^{1/2}/\sigma$, we find that the generating function scales as
\begin{align}
  g_a(z)&\sim - \frac{1}{4\pi^2(1-z)^2}\int_{-\infty}^\infty \int_{-\infty}^\infty dk_1 dk_2 \frac{1}{(k_1+k_2-i\epsilon)^2(1+k_1^2)(1+k_2^2)}+O\left[(1-z)^{-1}\right]\,,\quad z\rightarrow 1\,.\label{eq:gam}
\end{align}
Performing the integrals and letting $\epsilon \rightarrow 0$ gives
\begin{align}
  g_a(z)&\sim \frac{1}{16\,(1-z)^2}+O\left[(1-z)^{-1}\right]\,,\quad z\rightarrow 1\,.\label{eq:gal}
\end{align}
Using Tauberian theorem (\ref{eq:taub}) to invert the generating function (\ref{eq:ga}) gives
\begin{align}
  \sum_{m=0}^n \int_{0}^\infty dX\, X\,P(X,m)  P(X,n-m) \sim \frac{n}{16} + O(1)\,, \quad n\rightarrow \infty\,.\label{eq:numa}
\end{align}
Next, we find that the denominator in (\ref{eq:areap}) decays as [see formula (\ref{eq:denl}) for $\mu=2$ and $a=\sigma/\sqrt{2}$]
\begin{align}
  P(X=0,n) \sim \frac{1}{\sqrt{2\,\pi\,\sigma\,n}}+O(n^{-3/2})\,,\quad n\rightarrow \infty\,.\label{eq:px0}
\end{align}
Therefore, the expected absolute area under the bridge grows, when $n \to \infty$, as
\begin{align}
  \frac{1}{\sigma}\sum_{m=0}^n {\mathbb E}[|X_m|] \sim \frac{\sqrt{\pi}\,n^{3/2}}{2\,\sqrt{8}} + O(n^{1/2})\,,\label{eq:2area2}
\end{align}
where the correction is sublinear in $n$. Taking the difference between the expected area of the triangle linking the maximum and the bridge end points, and the expected absolute area under the bridge (see figure \ref{fig:bridge}), we find that the first order terms cancel exactly and what remains is the finite size constant, i.e., 
\begin{align}
 \lim_{n \to \infty}  \frac{\frac{n}{2}\,{\mathbb E}[M_n]-\sum_{m=0}^n {\mathbb E}[|X_m|]}{\sigma\,n} = \frac{\gamma}{2}\,,\label{eq:thermo2}
\end{align}
which is a signature that embodies the discreteness of the random walk.
\section{Applications}
\label{sec:app}
In this section, we discuss applications of our results to the convex hull of a tethered Rouse polymer chain (section \ref{sec:rouse}), the convex hull of a $2d$ bridge run-and-tumble particle (section \ref{sec:rtp})  and a bridge version of the lamb-lion problem (section \ref{sec:lamblion}).
\subsection{Tethered Rouse polymer chain}
\label{sec:rouse}
\begin{figure}[t]
  \begin{center}
    \includegraphics{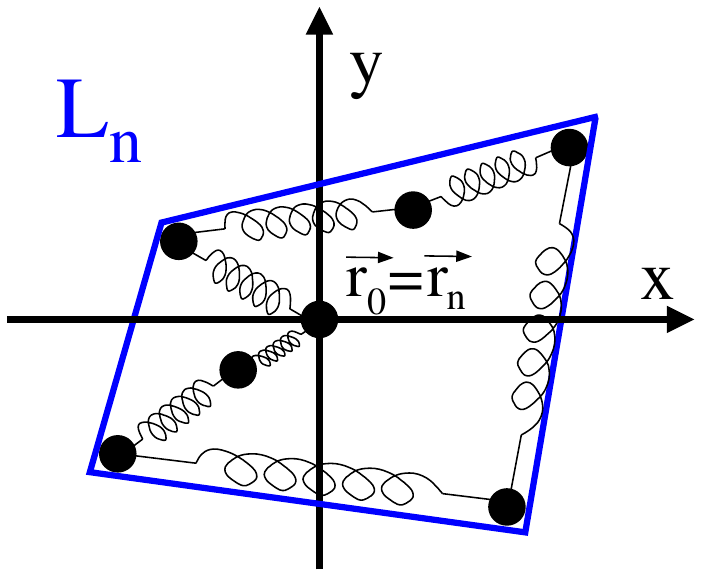}
    \caption{Tethered rouse polymer chain of $n=7$ beads. The chain is attached at both ends to the origin $\vec{r}_0=\vec{r}_n=\vec{0}$. The convex hull (blue line) has a perimeter $L_n$. }
    \label{fig:rouse}
  \end{center}
\end{figure}

Let us consider a $2d$ tethered Rouse polymer chain. The chain is made of $n$ beads located in the plane at $\{\vec{r}_i\}_n$ and are connected to each others by harmonics springs. The tethered polymer is attached to the origin at both ends $\vec{r}_0=\vec{r}_n=\vec{0}$. The probability of a given configuration $P[\{\vec{r}_i\}_n]$ is given by the Boltzmann distribution
\begin{align}
  P[\{\vec{r}_i\}_n] = \frac{1}{Z_n}\,\rme^{-\frac{\beta\,\kappa}{2}\,\sum_{i=1}^n(\vec{r}_i-\vec{r}_{i-1})^2}\,,\label{eq:prouse}
\end{align}
where $\beta=\frac{1}{k_B T}$ is the inverse temperature, $\kappa$ is the spring constant and $Z_n$ is the normalization constant. We are interested in the mean perimeter ${\mathbb E}[ L_n]$ of the convex hull of a Rouse polymer of length $n$ (see figure \ref{fig:rouse}). 
The Rouse polymer chain can be thought of a $2d$ bridge random walk
\begin{align}
  \vec{r}_{m+1} = \vec{r}_m + \vec{\eta}_m\,,\label{eq:eomR}
\end{align}
where $\vec{r}_0=\vec{r}_n=\vec{0}$ and $\vec{\eta}_m$ are i.i.d.~$2d$ Gaussian random variables distributed according to
\begin{align}
  f(\eta_x,\eta_y) = \frac{1}{2\,\pi\,\sigma^2}\,\rme^{-\frac{1}{2}\left(\frac{\eta_x^2+\eta_y^2}{\sigma^2}\right)}\,,\label{eq:Gdist2}
\end{align}
where $\sigma^2=\frac{1}{\beta\,\kappa}$.
To find the mean perimeter ${\mathbb E}[ L_n ]$ of the convex hull after $n$ jumps, we employ Cauchy formula \cite{Cauchy32}  (see e.g. \cite{Randon09Convex} for a derivation) which tells us that the mean perimeter is given by
\begin{align}
{\mathbb E}[L_n] = 2\pi\, {\mathbb E}[M_n]\,,\label{eq:Ln}
\end{align}
where ${\mathbb E}[M_n]$ is the expected maximum of the $x$ component of the $2d$ random walk (\ref{eq:eomR}) after $n$ jumps. The $x$ component of the $2d$ random walk (\ref{eq:eomR}) reduces to a Gaussian bridge random walk
\begin{align}
  x_{m+1} = x_m + \eta_m\,,\label{eq:eomR1}
\end{align} 
where $x_0=x_n$ and $\eta_m$ are i.i.d.~$1d$ Gaussian random variables distributed according to
\begin{align}
  f(\eta) =  \frac{1}{\sqrt{2\pi\,\sigma^2}}\,\rme^{-\frac{\eta^2}{2\sigma^2}}\,.\label{eq:rouseD1}
\end{align}
The Gaussian distribution being stable, the propagator of the free random walk (\ref{eq:eomR1}) is also a Gaussian distribution given by
\begin{align}
  P(X,m) = \frac{1}{\sqrt{2\pi\sigma^2\,m}}\,\rme^{-\frac{X^2}{2\sigma^2\,m}}\,.\label{eq:rouseProp1}
\end{align}
Inserting it into the exact expression of the expected maximum of a bridge random walk (\ref{eq:Mn}) and performing the integration, we find that expected maximum is given by
\begin{align}
  {\mathbb E}[M_n] = \frac{\sigma}{\sqrt{2\pi\,n}}\sum_{m=1}^{n-1}\, \sqrt{\frac{n-m}{m}}\,.\label{eq:rouseMn}
\end{align}
In the large $n$ limit, we find that it grows like
\begin{align}
{\mathbb E}[M_n] \sim  \sqrt{\frac{\sigma^2\,\pi\,n}{8}} +\sigma \,\frac{\zeta(1/2)}{\sqrt{2\pi}}\,,\quad n\rightarrow \infty\,,\label{eq:maxRMn}
\end{align}
where $\zeta(z)=\sum_{m=1}^{\infty} m^{-z}$ is the Riemann zeta function that has been analytically continued for $z<1$. The asymptotic limit (\ref{eq:maxRMn}) agrees with the general formula of phase I (\ref{eq:MasI}) as the jump distribution (\ref{eq:jumpRTP}) has a well defined finite variance and fourth moment. This can be seen using the following remarkable identity \cite{ComtetPrecise05} 
\begin{align}
\frac{1}{\pi\,\sqrt{2}}\int_0^\infty \frac{dk}{k^2}\,\ln\left(\frac{1-\rme^{-k^2}}{k^2}\right)=\frac{\zeta(1/2)}{\sqrt{2\pi}}.\label{eq:zeta}
\end{align}
Inserting the expression for the maximum (\ref{eq:maxRMn}) into Cauchy formula (\ref{eq:Ln}), we find that the expected perimeter of the convex hull of a tethered Rouse polymer chain grows like
\begin{align}
{\mathbb E}[L_n] =  \sqrt{\frac{\sigma^2\,\pi^2\,n}{2}} +\sigma \,\sqrt{2\pi}\,\zeta(1/2)\,,\quad n\rightarrow \infty\,.\label{eq:rouseLn}
\end{align}

\subsection{Bridge run-and-tumble particle in $d=2$}
\label{sec:rtp}
\begin{figure}[t]
  \begin{center}
    \includegraphics{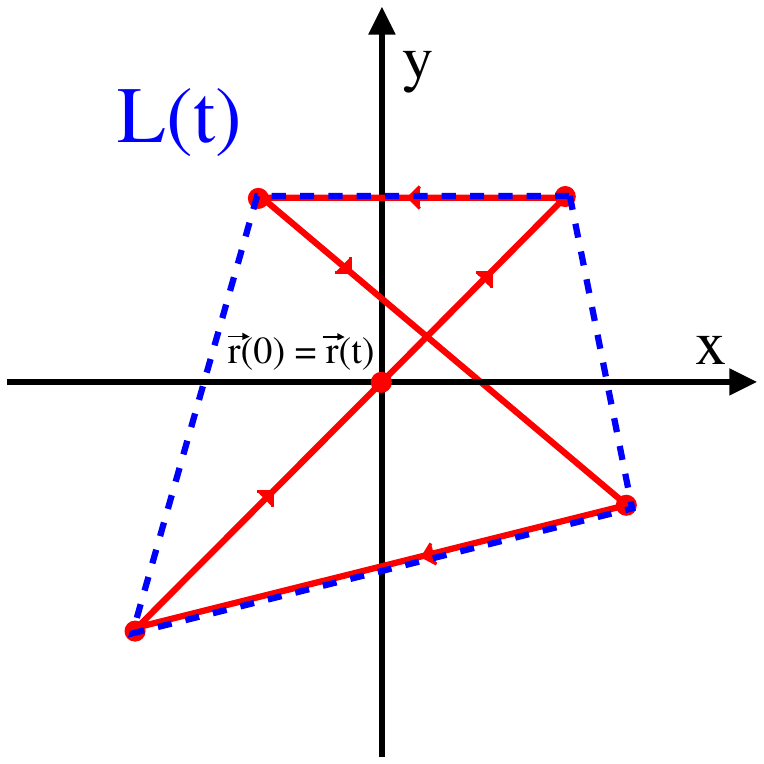}
    \caption[width=0.7\textwidth]{Bridge run-and-tumble trajectory in $2$ dimensions with $n=5$ tumbling events (red dots) during a time $t$. The particle is constrained to return to the origin at time $t$. The convex hull (blue line) has a perimeter $L(t)$. }
    \label{fig:rtp}
  \end{center}
\end{figure}

Let us consider a $2d$ bridge run-and-tumble particle with velocity $v_0$ and tumbling rate $\tilde \gamma$ (not to confuse with the constant $\gamma$ discussed in the introduction) that is constrained to return to the origin after a fixed time $t$. We quickly remind the reader what a run-and-tumble particle is. The particle starts at the origin with an initial velocity $v_0$ whose direction is drawn uniformly in $[0,2\pi[$. The particle then runs with such velocity during a time $\tau$ drawn from an exponential distribution 
\begin{align}
  p(\tau)=\tilde \gamma\, \rme^{-\tilde \gamma \tau}\,,\label{eq:exp}
\end{align} after which it tumbles and chooses again another direction uniformly in $[0,2\pi[$ and travels for another random time with a velocity $v_0$. The particle performs this motion \textit{ad infinitum}, hence the name ``run-and-tumble'' particle. It turns out that the mean perimeter of the convex hull of a \emph{free} run-and-tumble particle was obtained in \cite{HartmannConvex20}. In this section, we compute the mean perimeter of the convex hull of a \emph{bridge} run-and-tumble particle (see figure \ref{fig:rtp}) and compare it to the free result. We first obtain this result in the fixed $n$ ensemble where the total number of tumbling events $n$ are fixed. Then, we compute the mean perimeter in the fixed $t$ ensemble by summing over all possible number of events that can occur in a given time $t$. Note that in the following we count the origin as an initial tumble.

\subsubsection{Fixed $n$ ensemble:}

In the fixed $n$ ensemble, we view the run-and-tumble trajectory as a discrete $2d$ random walk whose jump distribution can be derived from the exponential run time distribution (\ref{eq:exp}). Indeed, in a time $\tau$, the particle will travel a distance $l=v_0\,\tau$. If we denote $\eta_x$ and $\eta_y$ the $2d$ components of the traveled distance such that $l=\sqrt{\eta_x^2+\eta_y^2}$, the jump distribution of the discrete random walk is given by
\begin{align}
  f(\eta_x,\eta_y) = \frac{\tilde \gamma}{v_0\sqrt{\eta_x^2+\eta_y^2}}\,\rme^{-\frac{\tilde \gamma}{v_0} \sqrt{\eta_x^2+\eta_y^2}}\,.\label{eq:rtp2df}
\end{align}
To use Cauchy formula (\ref{eq:Ln}), as in section \ref{sec:rouse}, we need to know the marginal distribution of $\eta_x$. Therefore, we integrate the $2d$ jump distribution (\ref{eq:rtp2df}) over the $\eta_y$ component and find that the marginal distribution of the $x$ component is given by \cite{HartmannConvex20}
\begin{align}
  f(\eta_x) = \frac{\tilde \gamma}{\pi\,v_0}\,K_0\left(\frac{\tilde \gamma\,|\eta_x|}{v_0}\right)\,,\label{eq:jumpRTP}
\end{align}
where $K_\nu(z)$ is the modified Bessel function of index $\nu$. The forward propagator after $n$ jumps of the $\eta_x$ component is given by \cite{HartmannConvex20}
\begin{align}
  P(X,m) =\int_{-\infty}^\infty \frac{dk}{2 \pi} \hat f(k)^m\,\rme^{-i\,k\,X}= \frac{\tilde \gamma\,2^{(1-m)/2} }{v_0\,\sqrt{\pi}\,\tilde \gamma(m/2)}\,\left(\frac{\tilde \gamma |X|}{v_0}\right)^{(m-1)/2}\, K_{(1-m)/2}\left(\frac{\tilde \gamma\,|X|}{v_0}\right)\,,\label{eq:Pm}
\end{align}
where $\hat f(k)$ is the Fourier transform of the jump distribution (\ref{eq:jumpRTP}).
Inserting the propagator (\ref{eq:Pm}) into the expression of the expected maximum of a bridge random walk (\ref{eq:Mn}) and integrating yields the exact expression
\begin{align}
   {\mathbb E}[M_n] &= \frac{v_0\,\Gamma
   \left(\frac{n}{2}\right) }{\tilde \gamma\sqrt{\pi}\,n\,\Gamma
   \left(\frac{n-1}{2}\right) }\sum_{m=1}^{n-1}\,\frac{ \Gamma
   \left(\frac{m+1}{2}\right)
   \Gamma \left(\frac{n-m+1}{2}
   \right)}{ \Gamma
   \left(\frac{m+2}{2}\right)
   \Gamma
   \left(\frac{n-m}{2}\right)}\,\,.\label{eq:avgM}
\end{align}
In the large $n$ limit, one obtains from the expression (\ref{eq:avgM}), that the expected maximum grows like
\begin{align}
   {\mathbb E}[M_n] &\sim \sqrt{\frac{\pi\,v_0^2\,n}{8\,\tilde \gamma^2}}-\frac{(2+\pi)\,v_0}{2\,\pi\,\tilde \gamma}\,,\quad n\rightarrow \infty\,,\label{eq:Mnlrtp}
\end{align}
which agrees with the general formula of phase I (\ref{eq:MasI}) as the jump distribution (\ref{eq:jumpRTP}) has a finite variance and fourth moment. Indeed, this can be seen by noting that the variance of the jump distribution (\ref{eq:jumpRTP}) is 
\begin{align}
  \sigma^2 = \int_{-\infty}^{\infty} d\eta_x \,\eta_x^2\, f(\eta_x) = \frac{v_0^2}{\tilde \gamma^2}\,,\label{eq:sigrtp}
\end{align}
and that the integral in the second leading term in (\ref{eq:MasI}) can be evaluated using the following identity
\begin{align}
  -\int_{-\infty}^{\infty} \frac{dk}{k^2}\,\ln\left[\frac{2}{k^2}\left(1-\frac{1}{\sqrt{1+k^2}}\right)\right] = 2+\pi \,.\label{eq:intrtp}
\end{align}
Using the relation in (\ref{eq:Ln}), the mean perimeter (\ref{eq:Ln}) of the convex hull is thus
\begin{align}
 {\mathbb E}[L_n] &= \frac{2\sqrt{\pi}\,v_0\,\Gamma
   \left(\frac{n}{2}\right) }{\tilde \gamma\,n\,\Gamma
   \left(\frac{n-1}{2}\right) }\sum_{m=1}^{n-1}\,\frac{ \Gamma
   \left(\frac{m+1}{2}\right)
   \Gamma \left(\frac{n-m+1}{2}
   \right)}{ \Gamma
   \left(\frac{m+2}{2}\right)
   \Gamma
   \left(\frac{n-m}{2}\right)}\,,\label{eq:Lnf}
   \end{align}
   which grows asymptotically as
   \begin{align}
  {\mathbb E}[L_n] &\sim  \sqrt{\frac{\pi^3\,v_0^2\,n}{2\,\tilde \gamma^2}}-\frac{(2+\pi)\,v_0}{\tilde \gamma}\,,\quad n\rightarrow \infty\,.\label{eq:Lnrtp}
\end{align}

\subsubsection{Fixed $t$ ensemble}
In a given time $t$, the probability that $n$ tumbling events occurred is given by the Poisson distribution
\begin{align}
  \text{Pr.}(N(t)=n) = \frac{\tilde \gamma^n\,t^n\,\rme^{-\tilde \gamma\,t}}{n!}\,.\label{eq:poisson}
\end{align}
At the $n^\text{th}$ jump, there has been $n-1$ tumbling events. Therefore, the mean perimeter as a function of time ${\mathbb E}[L(t)]$ can be obtained by summing over all possible number of tumbling events, i.e.,  
\begin{align}
  {\mathbb E}[L(t)]  &= \sum_{n=2}^\infty {\mathbb E}[L_n] \,\times  \text{Pr.}(N(t)=n-1) \,,\label{eq:ltln}\\
    &=\sum_{n=2}^\infty\frac{2\sqrt{\pi}\,v_0\,\Gamma
   \left(\frac{n}{2}\right) }{\tilde \gamma\,n\,\Gamma
   \left(\frac{n-1}{2}\right) }\sum_{m=1}^n\,\frac{ \Gamma
   \left(\frac{m+1}{2}\right)
   \Gamma \left(\frac{n-m+1}{2}
   \right)}{ \Gamma
   \left(\frac{m+2}{2}\right)
   \Gamma
   \left(\frac{n-m}{2}\right)}\times\frac{\tilde \gamma^{n-1}\,t^{n-1}\,\rme^{-\tilde \gamma\,t}}{(n-1)!}\,,\label{eq:Ltexact}
\end{align}
where the first sum starts at $n=2$ as it is the minimal bridge length.
    The sums in the exact expression of the mean perimeter (\ref{eq:Ltexact}) seem difficult to evaluate. Nevertheless, we can obtain the long time limit of the mean perimeter by replacing ${\mathbb E}[ L_n ]$ by its large $n$ approximation and the Poisson distribution $\text{Pr.}(N(t)=n)$ by a Gaussian distribution $\mathcal{N}(\mu=\tilde \gamma\,t,\sigma^2=\tilde \gamma\,t)$ to find
    \begin{align}
     {\mathbb E}[L(t)]  &\sim \int dn\,\left( \sqrt{\frac{\pi^3\,v_0^2\,n}{2\,\tilde \gamma^2}}-\frac{(2+\pi)\,v_0}{\tilde \gamma}\right)\, \frac{1}{\sqrt{2\pi \,\tilde \gamma\,t}}\,\rme^{-\frac{1}{2}\left(\frac{n-\tilde \gamma\,t}{\sqrt{\tilde \gamma\,t}}\right)^2}\,,\quad t\rightarrow \infty\,,\label{eq:LtG}\\
       &\sim \frac{v_0}{\tilde \gamma}\left[ \sqrt{\frac{\pi^3\,\tilde \gamma\,t}{2}}-(2+\pi)+O\left(t^{-\frac{1}{2}}\right)\right]\,,\quad t\rightarrow \infty\,.\label{eq:Ltlong}
    \end{align}
 Similarly, we can also obtain the small $t$ limit by only considering the first non-zero term in the sums in (\ref{eq:Ltexact}), which gives
    \begin{align}
    {\mathbb E}[L(t)]  &\sim \frac{2\,v_0\,t}{\pi}+O(t^2)\,,\quad t\rightarrow 0\,.\label{eq:Ltsm}
    \end{align}
  In summary the average perimeter of the convex hull behaves as
    \begin{align}
    {\mathbb E}[L(t)]  &\sim \left\{\begin{array}{ll}
        \frac{2\,v_0\,t}{\pi}+O(t^2)\,,& t\rightarrow 0\,,\\[1em]
        \frac{v_0}{\tilde \gamma}\left[ \sqrt{\frac{\pi^3\,\tilde \gamma\,t}{2}}-(2+\pi)+O\left(t^{-\frac{1}{2}}\right)\right]\,,& t\rightarrow \infty\,,
      \end{array}\right.\label{eq:Ltres}
    \end{align}
     which corresponds to a ballistic small time behavior and a diffusive long-time limit. This is a typical feature of run-and-tumble particles and was also observed in the perimeter $L^*(t)$ of the convex hull of \emph{free} run-and-tumble particles for which the average value grows like \cite{HartmannConvex20}
     \begin{align}
         {\mathbb E}[L^*(t)]  &\sim \left\{\begin{array}{ll}
        2\,v_0\,t+O(t^2)\,,& t\rightarrow 0\,,\\[1em]
        \frac{v_0}{\tilde \gamma}\left[ \sqrt{8\pi\tilde \gamma t}-(2+\pi)+O\left(t^{-\frac{1}{2}}\right)\right]\,,& t\rightarrow \infty\,.
      \end{array}\right.\label{eq:Lfree}
     \end{align}
  Comparing the leading first order term in (\ref{eq:Ltres}) and (\ref{eq:Lfree}), we see that the amplitude for the free run-and-tumble particle is larger than the one for the bridge one, which is expected as the bridge run-and-tumble particle cannot go as far as the free one due to its constraint to return to the origin.  Interestingly, the second order constant correction in the long-time limit is exactly the same as for the convex hull of free run-and-tumble particles.
   
\subsection{Lamb-lion problem}
\label{sec:lamblion}
\begin{figure}[t]
  \begin{center}
    \includegraphics[width=0.5\textwidth]{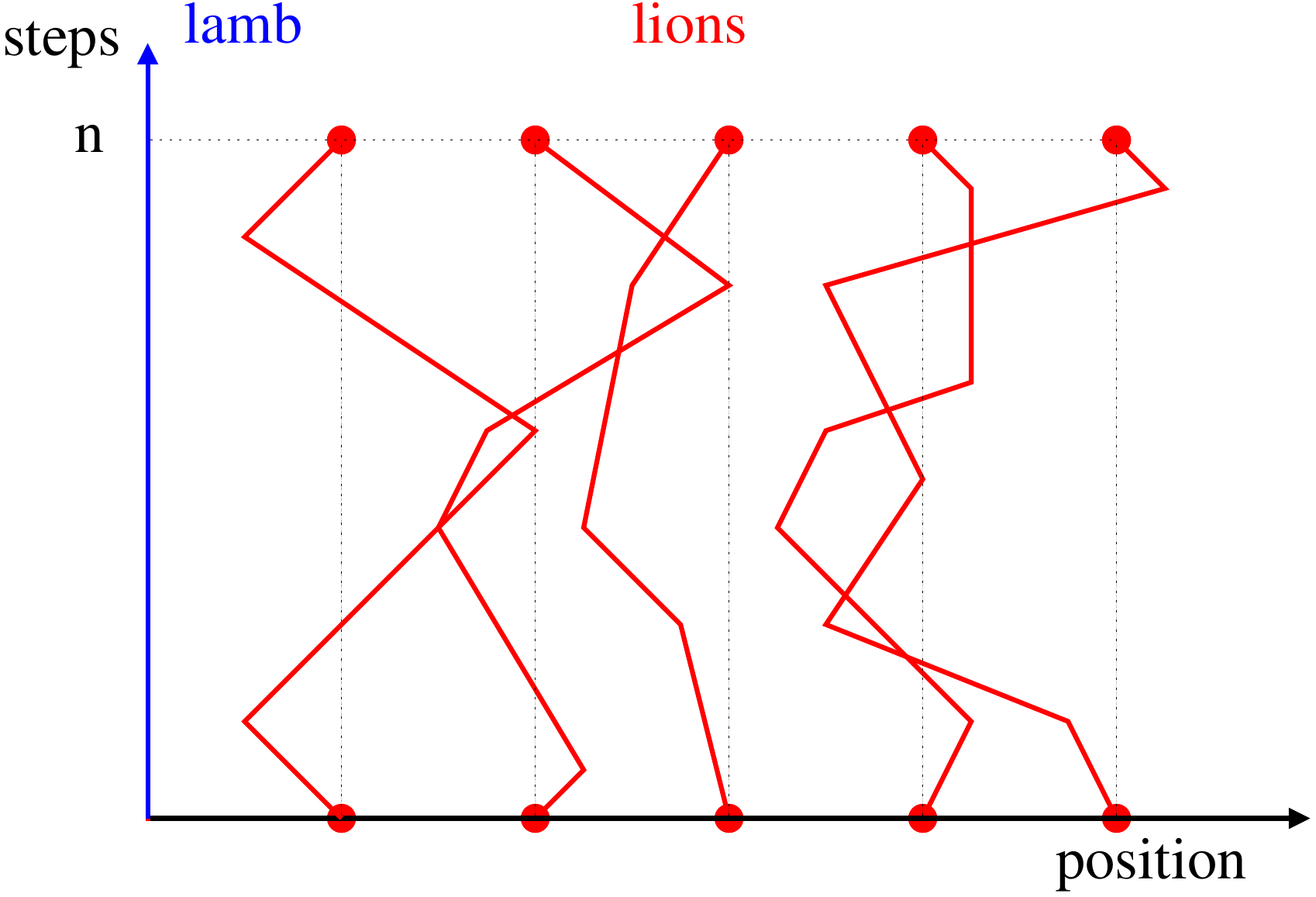}
    \caption{``Lamb-lion'' problem. The lamb is an immobile target located at the origin (blue). The lions, performing random walks, are initially uniformly distributed on the positive line with density $\rho_0$. The lions are further constrained to return to the origin at the $n^\text{th}$ step. The survival probability $S(n)$ is the probability that none of the lions have encountered the lamb during $n$ steps.}
    \label{fig:lamblion}
  \end{center}
\end{figure}

Finally, we consider a bridge version of the ``lamb-lion'' problem \cite{Franke12,Majumdar19Smol,Krapivsky96Lamb,Redner99Lamb,Redner01} where the lions are constrained to return to their initial position after their hunt. The setting is illustrated in figure \ref{fig:lamblion}. An immobile lamb is located at the origin and lions, performing random walks, are initially uniformly distributed on the positive line (with a uniform density $\rho_0$). 
Quite remarkably, the survival probability $S(n)$ that none of the lions have encountered the lamb is related to the expected maximum of a random walk by the following relation \cite{Majumdar19Smol}
\begin{align}
  S(n) = \exp(-\rho_0\, {\mathbb E}[M_n] )\,.\label{eq:Snll}
\end{align}
For large $n$, using our first order results on the expected maximum of a bridge random walk, we find that the survival probability of the lamb decays like
\begin{align}
S(n) \sim \exp(-\rho_0 \, a \, h_1(\mu) \, n^{\frac{1}{\mu}}) \quad, \quad n \to \infty \,, \label{eq:Snapprox}
\end{align}
where $h_1(\mu)$ is the amplitude given in (\ref{eq:h1intro}) and $a$ and $\mu$ are the parameters of the small-$k$ expansion of the jump distribution of the lions (\ref{eq:f}). Our results show that the second leading order correction to the expected maximum ${\mathbb E}[M_n]$ does not necessarily decay when $n$ is large. This means that the leading finite size correction therefore plays an important role as it will contribute to the amplitude of the decay of the survival probability. For instance, if the jump distribution of the lions is a Cauchy distribution with scale $\ell$ (\ref{eq:cauchyDist}), one needs to include the leading finite size correction to find that the survival probability decays as
\begin{align}
  S(n) \sim n^{\frac{1}{2\pi}\,\rho_0\,\ell}\,\rme^{-\frac{1}{8}\,\rho_0\,\ell\,\pi\,n}\,,\label{eq:survll}
\end{align}
up to a constant prefactor which would require an asymptotic analysis of the expected maximum up to third order to be determined. 

\section{Summary and outlook}
\label{sec:ccl}
In this work, we first obtained an explicit formula for the expected maximum of bridge discrete-time random walks of length $n$ with arbitrary jump distributions. This formula nicely extends an existing formula for free random walks. We then derived the asymptotic limit of the expected maximum for large $n$ up to second leading order and found a rich phase diagram depending on the jump distribution. In particular, we showed that, contrary to free random walks, bridge random walks with infinite first moment jump distributions have a well-defined expected maximum. We have also demonstrated how the leading finite size correction can appear in the large $n$ limit of a geometrical property of bridge random walks. Finally, we discussed applications of these results to the study of the convex hull of a tethered polymer chain, the convex hull of a $2d$ bridge run-and-tumble particle, and the survival probability in a bridge version of the lamb-lion problem.

Going beyond the expected value of the global maximum studied here, it would be interesting to investigate the full distribution of the global maximum as well as the order statistics, e.g.,  
the gap statistics between  consecutive maxima \cite{SM2012,LacroixGap19} of bridge random walks in the limit of large but finite $n$. Furthermore, it would be interesting to generalize our results to bridge random walks in higher dimensions, where one would for instance measure the maximum of the radial extent of the walk, and study how the leading finite size corrections are affected.

  \ack
  This work was partially supported by the Luxembourg National Research Fund (FNR) (App. ID 14548297).

\appendix

\section{Amplitude of the leading order term}
\label{app:h1}
In this appendix, we compute the double integrals in the amplitude $h_1(\mu)$ obtained in (\ref{eq:h1}). To do so, we start from (\ref{eq:h1}) and use that 
\begin{align}
  \frac{1}{(k_1+k_2-i\,\epsilon)^2} = -\int_0^\infty dy\,y\, e^{-i\,y(k_1+k_2-i \epsilon)}\,,\label{eq:h1a1}
\end{align}
which gives
\begin{align}
  h_1(\mu) = -\frac{1}{4\,\pi\,\Gamma\left(1+\frac{1}{\mu}\right)}\int_{-\infty}^\infty \int_{-\infty}^\infty dk_1 dk_2 \int_0^\infty dy\,y\, \frac{\ln(1+|k_1|^\mu)}{(1+|k_2|^\mu)}\, e^{-i\,y(k_1+k_2-i \epsilon)}\,.\label{eq:h1a2}
\end{align}
Then, we split the integral over the four quadrants of the ($k_1$-$k_2$) plane to get, using that $\cos(y(k_1+k_2))+\cos(y(k_1-k_2))=2\cos(yk_1)\cos(y k_2)$
\begin{align}
  h_1(\mu) = -\frac{1}{\pi\,\Gamma\left(1+\frac{1}{\mu}\right)}\int_{0}^\infty \int_{0}^\infty dk_1 dk_2 \int_0^\infty dy\, y\,\frac{\ln(1+|k_1|^\mu)}{(1+|k_2|^\mu)}\,\cos(yk_1)\cos(yk_2)\,e^{-\epsilon y}\,.\label{eq:h1a3} 
\end{align}
Finally, performing an integration by parts in the $k_1$ integral, one obtains 
\begin{align}
  h_1(\mu) = \frac{\mu}{\pi\,\Gamma\left(1+\frac{1}{\mu}\right)}\int_{0}^\infty  \int_{0}^\infty \int_0^{\infty}dy\,dk_1\,dk_2\,\frac{k_1^{\mu-1}\,\sin(yk_1)}{1+k_1^\mu}\,\frac{\cos(yk_2)}{1+k_2^\mu}e^{-\epsilon y} \,.\label{eq:h1a}
\end{align}
Using that $\sin(yk_1)\cos(yk_2)=[\sin(y(k_1+k_2))+\sin(y(k1-k2))]/2$, and integrating over $y$, we get
\begin{align}
   h_1(\mu) = \frac{\mu}{2\pi\,\Gamma\left(1+\frac{1}{\mu}\right)}\int_{0}^\infty  \int_{0}^\infty\,dk_1\,dk_2\,\frac{k_1^{\mu-1}}{1+k_1^\mu}\,\frac{1}{1+k_2^\mu} \left(\frac{1}{k_1+k_2}+\text{PV}\left\{\frac{1}{k_1-k_2}\right\}\right)\,,\label{eq:h1PV}
\end{align}
where we replaced the regularization parameter $\epsilon$ by the principal value (denoted PV). This principal value can be evaluated by anti symmetrizing the integral over $k_1$ and $k_2$, namely
\begin{align}
  h_1(\mu) = \frac{\mu}{2\pi\,\Gamma\left(1+\frac{1}{\mu}\right)}\int_{0}^\infty  \int_{0}^\infty\,dk_1\,dk_2\,\frac{1}{1+k_1^\mu}\,\frac{1}{1+k_2^\mu} \left(\frac{k_1^{\mu-1}}{k_1+k_2}+\frac{1}{2}\frac{k_1^{\mu-1}-k_2^{\mu-1}}{k_1-k_2}\right)\,.\label{eq:h1as}
\end{align}
The first term in the parenthesis can be symmetrized with respect to $k_1$ and $k_2$, which yields
\begin{align}
   h_1(\mu) = \frac{\mu}{4\pi\,\Gamma\left(1+\frac{1}{\mu}\right)}\int_{0}^\infty  \int_{0}^\infty\,dk_1\,dk_2\,\frac{1}{1+k_1^\mu}\,\frac{1}{1+k_2^\mu} \left(\frac{k_1^{\mu-1}+k_2^{\mu-1}}{k_1+k_2}+\frac{k_1^{\mu-1}-k_2^{\mu-1}}{k_1-k_2}\right)\,.\label{eq:h1s}
\end{align}
Simplifying the terms in the parenthesis, we get
\begin{align}
   h_1(\mu) = \frac{\mu}{2\pi\,\Gamma\left(1+\frac{1}{\mu}\right)}\int_{0}^\infty  \int_{0}^\infty\,dk_1\,dk_2\,\frac{k_1^{\mu }-k_2^{\mu }}{\left(k_1^2-k_2^2\right) \left(k_1^{\mu }+1\right)
   \left(k_2^{\mu }+1\right)}\,.\label{eq:h1symp}
\end{align}
In terms of polar coordinates $k_1=r\cos(\theta)$, $k_2=r\sin(\theta)$, it gives
\begin{align}
   h_1(\mu) = \frac{\mu}{2\pi\,\Gamma\left(1+\frac{1}{\mu}\right)}\int_{0}^\infty  \int_{0}^{\pi/2}\,dr\,d\theta\, \frac{r^{\mu-1}\left[\cos(\theta)^\mu-\sin(\theta)^\mu\right]}{\cos(2\theta)[1+r^\mu \cos(\theta)^\mu][1+r^\mu \sin(\theta)^\mu]}\,.\label{eq:h1pol}
\end{align}
Changing variable $u=r^\mu$, we get
\begin{align}
    h_1(\mu) = \frac{1}{2\pi\,\Gamma\left(1+\frac{1}{\mu}\right)}\int_{0}^\infty  \int_{0}^{\pi/2}\,du\,d\theta\, \frac{\left[\cos(\theta)^\mu-\sin(\theta)^\mu\right]}{\cos(2\theta)[1+u \cos(\theta)^\mu][1+u \sin(\theta)^\mu]}\,.\label{eq:h1polu}
\end{align}
Performing the integral over $u$, we find 
\begin{align}
    h_1(\mu) = \frac{\mu}{2\pi\,\Gamma\left(1+\frac{1}{\mu}\right)} \int_{0}^{\pi/2}\,d\theta\, \frac{\log(\cot(\theta))}{\cos(2\theta)}\label{eq:h1poli}\,.
\end{align}
Changing variables $v=\cot(\theta)$, we get
\begin{align}
   h_1(\mu) = \frac{\mu}{2\pi\,\Gamma\left(1+\frac{1}{\mu}\right)} \int_{0}^{\infty}\,dv\, \frac{\log(v)}{v^2-1}\,.\label{eq:h1v}
\end{align}
The last integral can be evaluated by integration by parts and gives $\pi^2/4$, hence we find the amplitude (\ref{eq:h1intro}) given in the introduction.

\section{Second leading order term of the expected maximum}
In this appendix, we derive the second leading order term of the expected maximum in the three phases presented in figure \ref{fig:phaseDiagram}.
\label{app:2max}
\subsection{Phase I: $1<\mu\leq 2$ and $\mu+1<\nu\leq 2 \mu$}
\label{app:case1}
We start from (\ref{eq:g2sc}) and write
\begin{align}
  \ln(1-z\hat f(\eta)) = \ln(1-z+a^\mu\,|\eta|^\mu)+ \ln\left(\frac{1-z\hat f(\eta)}{1-z+a^\mu\,|\eta|^\mu}\right)\,.\label{eq:caseIln}
\end{align}
Inserting this in the integral in (\ref{eq:g2sc}), we get 
\begin{align}
  g(z)&=g_1(z) + g_2(z)\,,\label{eq:gsumlev2}
\end{align}
where
\begin{align}
  g_1(z)&=\frac{(1-z)^{1/\mu}}{4\pi^2}\int_{-\infty}^\infty \int_{-\infty}^\infty dk_1 dk_2 \frac{\ln(1-z+a^\mu\,|k_1|^\mu)}{(k_1+k_2 (1-z)^{1/\mu}-i\epsilon)^2} \frac{1}{1-z\hat f(k_2(1-z)^{1/\mu})}\,,\label{eq:g1lev1}\\ 
g_2(z)&=\frac{(1-z)^{1/\mu}}{4\pi^2}\int_{-\infty}^\infty \int_{-\infty}^\infty dk_1 dk_2 \frac{\ln(\frac{1-z\hat f(k_1)}{1-z+a^\mu\,|k_1|^\mu})}{(k_1+k_2 (1-z)^{1/\mu}-i\epsilon)^2} \frac{1}{1-z\hat f(k_2(1-z)^{1/\mu})}\,.\label{eq:g2lev1}
\end{align}
\paragraph{Asymptotic limit of $g_1(z)$.} Rescaling $k_1$ by $(1-z)^{1/\mu}/a$ and $k_2$ by $a$, we find that $g_1(z)$ scales as
\begin{align}
  g_1(z)&\sim  \frac{1}{4\pi^2(1-z)}\int_{-\infty}^\infty \int_{-\infty}^\infty dk_1 dk_2 \frac{\ln(1+|k_1|^\mu)}{(k_1+k_2-i\epsilon)^2(1+|k_2|^\mu)}\,.\label{eq:g1simI}
\end{align}
\paragraph{Asymptotic limit of $g_2(z)$.} Letting $z\rightarrow 1$, we find that $g_2(z)$ scales as 
\begin{align}
  g_2(z)&\sim\frac{1}{4\pi^2\,(1-z)^{1-\frac{1}{\mu}}}\int_{-\infty}^\infty \int_{-\infty}^\infty dk_1 dk_2 \frac{\ln(\frac{1-\hat f(k_1)}{a^\mu\,|k_1|^\mu})}{(k_1+k_2 (1-z)^{1/\mu}-i\epsilon)^2} \frac{1}{1+a^\mu|k_2|^\mu}\,.\label{eq:g2simI}
\end{align}
The integral over $k_2$ gives
\begin{align}
  g_2(z)&\sim\frac{1}{2\pi\,a\,\mu\,\sin(\frac{\pi}{\mu})\,(1-z)^{1-\frac{1}{\mu}}}\int_{-\infty}^\infty  dk_1  \frac{\ln(\frac{1-\hat f(k_1)}{a^\mu\,|k_1|^\mu})}{k_1^2} \,.\label{eq:g2simI2}
\end{align}

\paragraph{Large $n$ limit of the numerator.}
To obtain the large $n$ limit of the numerator in (\ref{eq:Mn}), we use the Tauberian theorem (\ref{eq:taub}):
\begin{align}
  \sum_{m=1}^n \frac{1}{m}\int_{0}^\infty dy\, y\,P(y,m)  P(y,n-m)  &\sim\frac{1}{4\pi^2}\int_{-\infty}^\infty \int_{-\infty}^\infty dk_1 dk_2 \frac{\ln(1+|k_1|^\mu)}{(k_1+k_2-i\epsilon)^2(1+|k_2|^\mu)} \nonumber \\
  &\quad + \frac{n^{-\frac{1}{\mu}}}{2\pi\,a\,\mu\,\sin(\frac{\pi}{\mu})\,\Gamma\left(1-\frac{1}{\mu}\right)}\int_{-\infty}^\infty  dk_1  \frac{\ln(\frac{1-\hat f(k_1)}{a^\mu\,|k_1|^\mu})}{k_1^2} \,,\quad n\rightarrow\infty\,.\label{eq:numlll}
\end{align}

\paragraph{Large $n$ limit of the denominator.} To obtain the large $n$ limit of the denominator in (\ref{eq:Mn}), We perform the same jumps as in the main text (\ref{eq:denlm}) but keeping the next-to-leading order:
\begin{align}
P(y=0,n)&=\int_{-\infty}^\infty \frac{dk}{2\pi} \hat f(k)^n \nonumber\\&=  \int_{-\infty}^\infty \frac{dk}{2\pi} \rme^{ n \log(\hat f(k))} \,,\nonumber\\
   &\sim \int_{-\infty}^\infty \frac{dk}{2\pi} \rme^{ -n\left(a^\mu |k|^{\mu}-b |k|^{\nu}\right) }\,,\nonumber\\
   &\sim \frac{\Gamma \left(1+\frac{1}{\mu }\right)  n^{-\frac{1}{\mu} }}{\pi \,a }+\frac{b \,\Gamma \left(\frac{\nu +1}{\mu }\right) 
   n^{-\frac{\nu -\mu+1}{\mu }}}{\pi \mu\, a
   ^{\nu +1}}\,,\quad n\rightarrow\infty\,.\label{eq:denl}
   \end{align}
\paragraph{Large $n$ limit of the expected maximum.}
Combining the results (\ref{eq:numlll}) and (\ref{eq:denl}), we find
\begin{align}
 {\mathbb E}[M_n]   &\sim h_1(\mu)\,a\,n^{\frac{1}{\mu}}  + \frac{1}{2\pi}\int_{-\infty}^\infty  \frac{dk_1}{k_1^2} \ln\left(\frac{1-\hat f(k_1)}{a^\mu\,|k_1|^\mu}\right) \,,\quad n\rightarrow \infty\,, \label{eq:asMnlll}
\end{align}
where $h_1(\mu)$ is the amplitude given in (\ref{eq:h1}).

\subsection{Phase II: $0<\mu\leq 2$ and $\mu<\nu \leq 2\,\mu$ and  $\nu<\mu+1$}
\label{app:case2}
We start from (\ref{eq:g2sc}) and write
\begin{align}
  \ln(1-z\hat f(\eta)) = \ln(1-z+a^\mu\,|\eta|^\mu-b|\eta|^{\nu})+ \ln\left(\frac{1-z\hat f(\eta)}{1-z+a^\mu\,|\eta|^\mu-b|\eta|^{\nu}}\right)\,.\label{eq:lnII}
\end{align}
Inserting this in the integral in (\ref{eq:g2sc}), we get 
\begin{align}
  g(z)&=g_1(z) + g_2(z)\,,\label{eq:gsumlev}
\end{align}
where
\begin{align}
  g_1(z)&=\frac{(1-z)^{\frac{1}{\mu}}}{4\pi^2}\int_{-\infty}^\infty \int_{-\infty}^\infty dk_1 dk_2 \frac{\ln(1-z+a^\mu\,|k_1|^\mu-b|k_1|^{\nu})}{(k_1+k_2 (1-z)^{1/\mu}-i\epsilon)^2} \frac{1}{1-z\hat f(k_2(1-z)^{1/\mu})}\,,\label{eq:g1lev}\\ 
g_2(z)&=\frac{(1-z)^{\frac{1}{\mu}}}{4\pi^2}\int_{-\infty}^\infty \int_{-\infty}^\infty dk_1 dk_2 \frac{\ln(\frac{1-z\hat f(k_1)}{1-z+a^\mu\,|k_1|^\mu-b|k_1|^{\nu}})}{(k_1+k_2 (1-z)^{1/\mu}-i\epsilon)^2} \frac{1}{1-z\hat f(k_2(1-z)^{1/\mu})}\,.\label{eq:g2lev}
\end{align}
One can show that the leading terms come from $g_1(z)$ as $g_2(z)$ will lead to corrections to ${\mathbb E}[M_n]$ that vanish when $n\rightarrow\infty$.
\paragraph{Asymptotic limit of $g_1(z)$.} Rescaling $k_1$ by $(1-z)^{1/\mu}/a$ and $k_2$ by $a$, we find that $g_1(z)$ scales as
\begin{align}
  g_1(z)&\sim  \frac{1}{4\pi^2(1-z)}\int_{-\infty}^\infty \int_{-\infty}^\infty dk_1 dk_2 \frac{\ln(1+|k_1|^\mu-b\,a^{-\frac{\nu}{\mu}}\,|k_1|^{\nu}(1-z)^{\frac{\nu-\mu}{\mu}})}{(k_1+k_2-i\epsilon)^2[1+|k_2|^\mu-\frac{b}{a^{\nu}}|k_2|^{\nu}(1-z)^{\frac{\nu-\mu}{\mu}}]}\,.\label{eq:g1ln12}
\end{align}
 We expand (\ref{eq:g1ln12}) in the limit $z\rightarrow 1$:
\begin{align}
   g_1(z)&\sim  \frac{1}{4\pi^2(1-z)}\int_{-\infty}^\infty \int_{-\infty}^\infty dk_1 dk_2 \frac{\ln(1+|k_1|^\mu)}{(k_1+k_2-i\epsilon)^2(1+|k_2|^\mu)}\,\nonumber \\
  & +\frac{b}{4\pi^2\,a^{\nu}\,(1-z)^{-\frac{\nu}{\mu}}}\int_{-\infty}^\infty \int_{-\infty}^\infty dk_1 dk_2 \frac{\ln(1+|k_1|^\mu)\,|k_2|^{\nu}}{(k_1+k_2-i\epsilon)^2(1+|k_2|^\mu)^2}\,\nonumber \\
   & -  \frac{b}{4\pi^2\,a^{\nu}\,(1-z)^{-\frac{\nu}{\mu}}}\int_{-\infty}^\infty \int_{-\infty}^\infty dk_1 dk_2 \frac{|k_1|^{\nu}}{(k_1+k_2-i\epsilon)^2(1+|k_2|^\mu)\,(1+|k_1|^\mu)}\,.\label{eq:g1limII}
\end{align}
\paragraph{Large $n$ limit of the numerator.}
To obtain the large $n$ limit of the numerator in (\ref{eq:Mn}), we use the Tauberian theorem (\ref{eq:taub}). We obtain
\begin{align}
  \sum_{m=1}^n \frac{1}{m}\int_{0}^\infty& dy\, y\,P(y,m)  P(y,n-m)   \sim \nonumber\\
  &\frac{1}{4\pi^2}\int_{-\infty}^\infty \int_{-\infty}^\infty dk_1 dk_2 \frac{\ln(1+|k_1|^\mu)}{(k_1+k_2-i\epsilon)^2(1+|k_2|^\mu)}\,\nonumber \\
   & -  \frac{b\,n^{-\frac{\nu-\mu}{\mu}}}{4\pi^2\,a^{\nu}\,\Gamma\left(-\frac{\nu}{\mu}\right)}\int_{-\infty}^\infty \int_{-\infty}^\infty dk_1 dk_2 \frac{|k_1|^{\nu}}{(k_1+k_2-i\epsilon)^2(1+|k_2|^\mu)\,(1+|k_1|^\mu)}\nonumber\\
   & + \frac{b\,n^{-\frac{\nu-\mu}{\mu}}}{4\pi^2\,a^{\nu}\,\Gamma\left(-\frac{\nu}{\mu}\right)}\int_{-\infty}^\infty \int_{-\infty}^\infty dk_1 dk_2 \frac{\ln(1+|k_1|^\mu)\,|k_2|^{\nu}}{(k_1+k_2-i\epsilon)^2(1+|k_2|^\mu)^2}
   \,.\label{eq:numllln}
\end{align}

\paragraph{Large $n$ limit of the expected maximum.}
Combining the results (\ref{eq:denl}) and (\ref{eq:numllln}) we find
\begin{align}
  {\mathbb E}[ M_n] &\sim h_1(\mu)\,a\,n^{\frac{1}{\mu}} 
  + \frac{b}{a^{\nu-1}} \,h_2(\mu,\nu)\,n^{\frac{1-\nu+\mu}{\mu}}\,,\label{eq:mnIapp}
\end{align}
where $h_1(\mu)$ is given in (\ref{eq:h1}) and $h_2(\mu,\nu)$ is given by
\begin{align}
  h_2(\mu,\nu)&= \frac{1}{4\pi\,\Gamma\left(1+\frac{1}{\mu}\right)}\,\Bigg[\frac{1}{\Gamma\left(-\frac{\nu}{\mu}\right)}\int_{-\infty}^\infty \int_{-\infty}^\infty dk_1 dk_2 \frac{\ln(1+|k_1|^\mu)\,|k_2|^{\nu}}{(k_1+k_2-i\epsilon)^2(1+|k_2|^\mu)^2}\nonumber\\
  & -\frac{1}{\Gamma\left(-\frac{\nu}{\mu}\right)}\int_{-\infty}^\infty \int_{-\infty}^\infty dk_1 dk_2 \frac{|k_1|^{\nu}}{(k_1+k_2-i\epsilon)^2(1+|k_2|^\mu)\,(1+|k_1|^\mu)}\nonumber\\
  -&\frac{\Gamma\left(\frac{\nu+1}{\mu}\right)}{\mu\,\Gamma\left(1+\frac{1}{\mu}\right)}\int_{-\infty}^\infty \int_{-\infty}^\infty dk_1 dk_2 \frac{\ln(1+|k_1|^\mu)}{(k_1+k_2-i\epsilon)^2(1+|k_2|^\mu)}\Bigg]\,. \label{eq:h2ne2}
\end{align}
Performing similar steps to the ones done in \ref{app:h1}, one can simplify the double integration and obtain the result displayed in the introduction (\ref{eq:h2s}).

\subsection{Phase III: $1<\mu\leq 2$ and $\nu=\mu+1$}
\label{app:case3}
As in case II, only $g_1(z)$ will contribute to the two first order terms in $g(z)$ except that now the second leading order term in logarithmic. For $\nu=2$, $g_1(z)$ in (\ref{eq:g1lev}) is given by
\begin{align}
   g_1(z)&=\frac{1}{4\pi^2(1-z)}\int_{-\infty}^\infty \int_{-\infty}^\infty dk_1 dk_2 \frac{\ln(1+|k_1|^\mu+ c\,a^{-\mu-1}\,|k_1|^{\mu+1}(1-z)^{\frac{1}{\mu}})}{(k_1+k_2-i\epsilon)^2(1+|k_2|^\mu)}\,.\label{eq:g1l1}
\end{align}
We expand the logarithm in the integral  (\ref{eq:g1l1}) but only in the integrable region and we get
\begin{align}
 g_1(z)&\sim   \frac{1}{4\pi^2(1-z)}\int_{-\infty}^\infty \int_{-\infty}^\infty dk_1 dk_2 \frac{\ln(1+|k_1|^\mu)}{(k_1+k_2-i\epsilon)^2(1+|k_2|^\mu)}\,\nonumber \\
   & -  \frac{b}{4\pi^2\,a^{\mu+1}\,(1-z)^{1-\frac{1}{\mu}}}\int_{|k_1|<\frac{1}{(1-z)^{\frac{1}{\mu}}}} \int_{-\infty}^\infty dk_1 dk_2 \frac{|k_1|^{\mu+1}}{(k_1+k_2-i\epsilon)^2(1+|k_2|^\mu)\,(1+|k_1|^\mu)}\,.\label{eq:g1III}
\end{align}
Performing the integration in the second term and letting $\epsilon\rightarrow 0$ gives
  \begin{align}
   g_1(z)&\sim  \frac{1}{4\pi^2(1-z)}\int_{-\infty}^\infty \int_{-\infty}^\infty dk_1 dk_2 \frac{\ln(1+|k_1|^\mu)}{(k_1+k_2-i\epsilon)^2(1+|k_2|^\mu)}\,\nonumber \\
   & -\frac{b}{\pi\mu^2\sin(\frac{\pi}{\mu})\,a^{\mu+1}}\,\ln\left(\frac{1}{1-z}\right)\frac{1}{(1-z)^{1-\frac{1}{\mu}}}\,,\quad z\rightarrow 1\,.\label{eq:g1n1}
\end{align}

\paragraph{Large $n$ limit of the numerator.}
To obtain the large $n$ limit of the numerator in (\ref{eq:Mn}), we use the Tauberian theorem (\ref{eq:taub}). We obtain
\begin{align}
  \sum_{m=1}^n \frac{1}{m}\int_{0}^\infty dy\, y\,P(y,m)  P(y,n-m)  &\sim \frac{1}{4\pi^2}\int_{-\infty}^\infty \int_{-\infty}^\infty dk_1 dk_2 \frac{\ln(1+|k_1|^\mu)}{(k_1+k_2-i\epsilon)^2(1+|k_2|^\mu)}\,\nonumber \\
   & -\frac{b}{\pi\mu^2\sin(\frac{\pi}{\mu})\,a^{\mu+1}\,\Gamma\left(1-\frac{1}{\mu}\right)}\,\frac{\ln(n)}{n^{\frac{1}{\mu}}}\,,\quad n\rightarrow \infty\,.\label{eq:numll1n}
\end{align}
\paragraph{Large $n$ limit of the expected maximum.}
Combining the results (\ref{eq:numll1n}) and (\ref{eq:denl}), we find
\begin{align}
  {\mathbb E}[ M_n] &\sim h_1(\mu)\,a\,n^{\frac{1}{\mu}} 
  -\frac{b}{a^\mu\,\pi\,\mu}\ln(n)\,,\label{eq:MnIIIapp}
\end{align}
where $h_1(\mu)$ is given in (\ref{eq:h1}).

\end{document}